\documentclass[12pt]{article}
\usepackage{epsfig}
\usepackage{amssymb}
\usepackage[english]{babel}
\usepackage{rotate}
\usepackage{graphicx}
\usepackage{amssymb,epsfig}
\usepackage{color}

\renewcommand{\theequation}{\arabic{section}.\arabic{equation}}

\newcommand{\beq}[1]{
\begin{equation}\label{#1}}
\newcommand{\eeq}{\end{equation}}
\newcommand{\bea}[1]{
\begin{eqnarray}\label{#1}}
\newcommand{\eea}{\end{eqnarray}}

\newcommand{\bra}[1]{\left\langle #1 \right|}
\newcommand{\ket}[1]{\left| #1 \right\rangle}

\newcommand{\Gl}[1]{Eq.~(\ref{#1})}

\newcommand{\al}{\alpha}
\newcommand{\be}{\beta}  
\newcommand{\ep}{\varepsilon}
\newcommand{\ga}{\gamma}

\newcommand{\la}{\lambda}

\newcommand{\si}{\sigma}
\newcommand{\ro}{\varrho}
\newcommand{\dd}{{\rm d}}

\newcommand{\pslash}{\!\not\!{P}}

\newcommand{\nn}{\nonumber}

\newcommand{\lash}[1]{\not\! #1 \,}

\newcommand{\wwV}{\widetilde{\widetilde{V}}}
\newcommand{\wwT}{\widetilde{\widetilde{T}}}

\newcommand{\s}{\hspace{-0.21cm}\slash}

\begin{document}

%%%%%%%%%%% Titlepage

\begin{titlepage}

\vspace*{1cm}

%\vspace*{-2cm}
%\begin{flushright}
%\begin{tabular}{l}
%TPR-00-10\\
%hep-ph/0510237
%\end{tabular}
%\end{flushright}
%\vskip2.5cm

\begin{center}
{\Large \bf Light Cone Sum Rules for $\gamma^*N \to \Delta$ Transition Form Factors}
\vspace{1cm}
\end{center}
\centerline{\large\sc V.M.~Braun$^a$, A.~Lenz$^a$, G.~Peters$^a$ and  
A.V.~Radyushkin$^{b,c}$\footnote{Also at Laboratory of Theoretical Physics, JINR, Dubna, Russia}}  
\vspace{1 cm}  
\centerline{$^a${\em  Institut f{\"u}r Theoretische Physik,  
Universit{\"a}t Regensburg,  
D-93040 Regensburg, Germany}}  
\centerline{$^b${\it  Physics Department, Old Dominion University,  
Norfolk, VA 23529, USA}}
\centerline{$^c${\it Theory Group, Jefferson Laboratory, Newport News, VA 23606, USA}}  
\vspace{1cm}  
\bigskip  
%\centerline{\large \em \today}  
\bigskip  
\vfill
\begin{center}
  {\large\bf Abstract\\[10pt]} \parbox[t]{\textwidth}{
A theoretical framework is suggested for the calculation of $\gamma^*N \to \Delta$ 
 transition form factors using the light-cone sum rule approach. 
Leading-order sum rules are derived and compared with the existing experimental 
data. We find that the transition form factors in a several GeV region are dominated 
by the ``soft'' contributions that can be thought of as overlap integrals of the 
valence components of the hadron wave functions.
The ``minus'' components of the quark fields contribute significantly 
to the result, which can be reinterpreted as large contributions of the 
quark orbital angular momentum.} 
\end{center}

\vspace*{1cm}

\\

\vspace*{\fill}
\eject
\end{titlepage}

%\documentclass[12pt]{book}
%\usepackage{german,color}
%\begin{document}
%\small
%\newcommand{\s}{\hspace{-0,21cm}\slash}
\tableofcontents
\newpage

%%%%%%%%%%%%%%%%%%%%%%%%%%%%
\setcounter{equation}{0} \section{Introduction}
%%%%%%%%%%%%%%%%%%%%%%%%%%%%

The concept of form factors plays an extremely important role in the
studies of the internal structure of composite particles.
The non-trivial dependence of  form factors on the  momentum transfer $Q^2$ 
(i.e.,  its deviation from the constant behavior) is usually 
a signal of  the non-elementary nature of the  investigated particle.
In particular, the pioneering study  of the nucleon form factors by Hofstadter
and collaborators 
\cite{Mcallister:1956ng} 
demonstrated that the nucleons have a finite size 
of the order of a fermi.  Later, it was observed 
 that the behavior of the proton electromagnetic form factors,
  in a rather wide range of momentum transfers,  
is well described by the so-called dipole formula 
$G_p(Q^2)/G_p(0) \approx G_D (Q^2) \equiv 1/(1+Q^2/0.71{\rm GeV}^2)^2$,
suggesting a simple $G_p(Q^2) \sim 1/(Q^2)^2$ power law for
their large-$Q^2$ asymptotic behavior.  
At the same time,  strong evidence was accumulated that 
the pion electromagnetic form factor is well described by the 
$\rho$-pole fit $F_{\pi} (Q^2) \approx   1/(1+Q^2/m_{\rho}^2)$ indicating that,
in the pion case, the asymptotic behavior looks more like  $1/Q^2$.
From the quark model point of view, the faster decrease 
of the proton form factor  seems rather natural, since 
the  proton contains more valence constituents 
than the pion. Furthermore, it was established that,
if one can treat the hadrons at high momentum transfer
as collinear beams of $N$ valence quarks located  at small transverse  
separations and exchanging  intermediate 
gluing particles with which  they interact via 
a dimensionless coupling constant,
then the spin-averaged form factor  behaves asymptotically as $1/(Q^2)^{N-1}$
\cite{Brodsky:1973kr}.  
This  hard-exchange picture and the resulting 
  dimensional power  counting rules \cite{Brodsky:1973kr,Matveev:1973ra}
can be  formally  extended onto other hard exclusive processes.

After the advent of quantum chromodynamics, 
this hard-gluon-exchange picture was 
formalized with  the help of the
 QCD factorization approach  to  exclusive processes 
\cite{Chernyak:1977as,Radyushkin:1977gp,Lepage:1979zb}
that presents one of the highlights of perturbative QCD (pQCD).
Within this approach, the hard gluon exchange contribution proves to  
be dominant for sufficiently large momentum transfers $Q^2$.
An important ingredient of the asymptotic pQCD formalism for 
 hard exclusive processes is the concept of hadron distribution amplitudes (DAs). 
They are  fundamental nonperturbative  
functions describing the momentum  distribution  within rare parton configurations 
when the hadron is represented by   
a fixed  number of Fock constituents.  
It was shown that in the $Q^2 \to \infty$ limit,
form factors can be  written  in a factorized form, as a convolution 
of distribution amplitudes related to  hadrons in the initial and  final state  
times a ``short-distance'' coefficient function that 
is calculable in QCD perturbation theory.
The leading contribution corresponds to DAs with minimal possible number 
of constituents, e.g., 3 for the proton and 2 for the pion.

The essential requirement for the 
applicability  of the pQCD approach is a high virtuality 
of the exchanged gluons and also of the quarks inside the 
short distance subprocess.
Since the quarks carry only some fractions $x_iP$, $y_j P'$ of the initial $P$
and final  $P'$ momenta, the virtualities of the internal lines
of the subprocess are generically given by $x_i y_j Q^2$, i.e., they
may be essentially smaller than $Q^2$, the nominal 
momentum transfer to the hadron. Assuming that $\langle x \rangle \sim 1/N$,
one should expect the reduction factor of $0.1$ for the proton and 
  $0.2$ for the pion. In the pion case, this expectation 
  was confirmed by an explicit calculation of the one-loop pQCD 
  radiative corrections
  \cite{Dittes:1981aw}. Absorbing the   terms proportional
  to the $\beta$-function coefficient $\beta_0$ into the effective  coupling constant 
  $\alpha_s(\mu^2)$ 
 of the hard gluon exchange, one indeed obtains  
  $\langle x \rangle \langle y \rangle Q^2$  as its argument. 
  As a result, at accessible $Q^2$,   the  bulk part of the hard pQCD contribution 
 comes from the regions where the ``hard'' virtualities are 
 much smaller than  the  typical hadronic scale of 1\,GeV$^2$ 
 \cite{Efremov:1980mb,Isgur:1988iw,Radyushkin:1990te}.  
 According to the pQCD factorization recipe, 
 contributions from such regions should not be included into  the hard term, 
 which is  strongly reduced after such contributions are subtracted. In  practice,
 the  subtraction is never made, and   pQCD estimates  are 
 based on the original   expressions (which implies, in particular,
 that the perturbative $\sim 1/k^2$ behavior of propagators
 is trusted even if $k^2 \to 0$). 
 Despite  this, in most cases pQCD results for hadronic form factors 
 need special efforts to bring their magnitude  close to experimental data.
 For example, assuming the ``asymptotic'' form $\varphi_{\pi} (x) =6 f_{\pi} x_1 x_2$  
 for the pion DA  gives $Q^2 F^{\rm as}_{\pi}(Q^2)  =
 8\pi f_{\pi}^2 \alpha_s \approx \alpha_s \times 0.44\,$GeV$^2$ 
 for the pion form factor \cite{Radyushkin:1977gp,Lepage:1979zb}
 that agrees with existing data only if one takes  an  uncomfortably 
 large value  $\alpha_s \approx 1$ for the ``hard'' gluon vertex.
 Switching  
 to a wider Chernyak-Zhitnitsky (CZ) shape  
 $\varphi_{\pi}^{CZ} (x) =30 f_{\pi} x_1 x_2 (x_1-x_2)^2$  
 \cite{Chernyak:1981zz,Chernyak:1983ej} 
 gives  $Q^2 F^{\rm CZ}_{\pi}(Q^2) = \frac{200}{9} \pi f_{\pi}^2 \alpha_s$,
 which formally agrees  with the data for  $\alpha_s \approx 0.4$.
 However, at accessible $Q^2$ more than 90\% of this contribution comes 
 from the region of  gluon virtualities below (500\,MeV)$^2$ \cite{Isgur:1988iw}.
 
 In the nucleon case, the situation is even worse.  
 For the asymptotic $\sim x_1 x_2 x_3$  form of the leading twist three-quark  
distribution amplitude, the proton form factor $G^p_M$ turns out to be zero 
to leading order \cite{Lepage:1979za,Avdeenko}, while the neutron form factor $G_M^n$
is of opposite sign compared to the data.
 Assuming the equal sharing DA $\sim \delta (x_1-1/3) \delta (x_2-1/3) \delta (x_3-1/3)$
 gives wrong signs both for the  proton and neutron  form factors
 \cite{Aznaurian:1979zz}.  
 Furthermore, if one takes the QCD sum rule estimate
 for the $\langle qqq\,|\, N \rangle $ matrix element, the absolute magnitude 
 of the form factors in the above examples is too small (by a factor of hundred) 
 \cite{Belyaev:1982sa} compared to the data.  
Just like in the pion case,  the  magnitude of the formal  pQCD result can be 
increased, and also the signs of the predicted proton and neutron magnetic 
form factors reversed to coincide with the experimental ones,  
by using CZ-type DAs \cite{Chernyak:1984bm,King:1986wi,Gari:1986dr,Chernyak:1987nu} 
having peaks in a region where the 
momentum fraction  of one of the quarks is close to 1.
Since  the average fractions of the nucleon momentum carried by the two other
quarks are small, this formal  result is  strongly dominated  for 
accessible $Q^2$ by regions 
of unacceptably small virtualities. 
It was  argued \cite{Li:1992nu} that  higher order Sudakov-type corrections
squeeze the size of the valence quark configuration 
participating in the  pQCD subprocess. Indeed, in the pion case, there are negative 
terms in the one-loop correction to the short-distance 
amplitude  that 
can be written as Sudakov double logarithms in the impact parameter
$b$-space \cite{Musatov:1997pu}. 
After  resummation  to all orders, they produce a factor 
like $\exp[-\alpha_s \ln^2 (Q^2b^2)/3 \pi]$  suppressing the contribution 
of large transverse separations. These effects
increase the region of the $x,y$  fractions
where the leading-order pQCD expressions are formally  applicable,
though they are not strong enough to visibly suppress nonperturbative regions
for accessible momentum transfers,
see e.g. \cite{Bolz:1994hb} for  discussion of the nucleon form factor case. 

\begin{figure}[ht]
\centerline{\epsfxsize15cm\epsffile{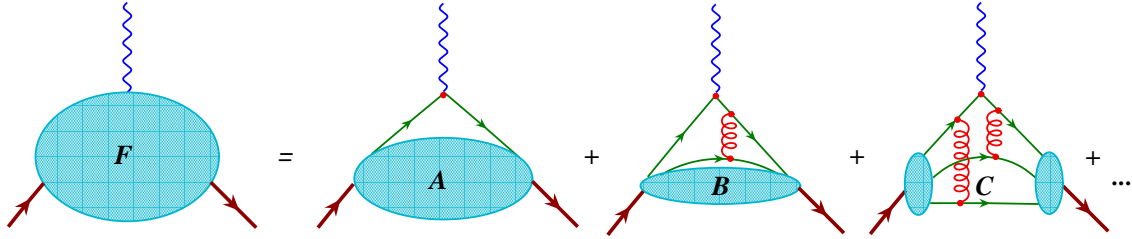}}
\caption{\label{figwan}\small
Structure of QCD factorization for baryon form factors.
}
\end{figure}

As already emphasized, according to the standard  philosophy of separating large- 
and small-virtuality  contributions underlying the pQCD factorization formulas,
the low-virtuality contributions  of 
gluon-exchange diagrams should be treated as a part of the 
soft contribution. More  precisely, in the case of the nucleon form factors,
the hard pQCD contribution  is only the third term of the  factorization 
expansion. Schematically, one can envisage the expansion of, 
say, the Dirac electromagnetic nucleon form factor
$F_1(Q^2)$ of the form (see Fig.~\ref{figwan})
\beq{schema}
F_1(Q^2) \sim A(Q^2)
+ \left ( \frac{\alpha_s(Q^2)}{\pi}\right )  \frac{B(Q^2)}{Q^2} 
+ \left ( \frac{\alpha_s(Q^2)}{\pi}\right ) ^2  \frac{C}{Q^4} + 
\ldots 
\eeq
where $C$ is  a constant determined by the nucleon DAs, while $A(Q^2)$ and  $B(Q^2)$
are  form-factor-type functions generated by soft contributions.  
Because of their   nonperturbative nature, it is impossible 
to tell precisely what is their large-$Q^2$ behavior.   On general grounds, one may  expect
that $A(Q^2)$  and $B(Q^2)/Q^2$  correspond to higher powers
of $1/Q^2$ than   the perturbative $1/Q^4$  term.
  Perturbative estimates suggest that
 $ A(Q^2), B(Q^2)/Q^2 \lesssim 1/Q^6$. 
 At  very large $Q^2$, one may also expect that they are further  suppressed by 
 Sudakov form factor.
The most important feature of the factorization expansion is a numerical suppression
of each hard gluon exchange by the $\alpha_s/\pi$ factor, which is a standard perturbation theory 
penalty for each extra loop. 
If 
$\alpha_s \sim 0.3$,    
the pQCD 
contribution to baryon form factors is suppressed by a factor of 
100 compared to the purely soft term.  
Thus,  one may expect that 
the onset of the perturbative regime is postponed to very large momentum transfers since 
the factorizable pQCD contribution ${ O}(1/Q^4)$ has to win over nonperturbative effects 
that are suppressed by  extra powers of $1/Q^2$, but do not involve small coefficients.     
In the light cone formalism, the functions  like $A(Q^2)$ and $B(Q^2)$ in the above expansion are determined by 
overlap integrals of the soft parts of  hadronic wave functions corresponding to large 
transverse separations. There is a growing consensus that such ``soft'' 
contributions play the dominant role at present energies. 
Indeed, it is known for a long time that the use of 
QCD-motivated models for the  wave functions allows one to
obtain, without much effort,  soft contributions comparable in size 
to  experimentally observed values (see, e.g.~\cite{Isgur:1984jm,Isgur:1988iw,Kroll:1995pv}). 
A new trend \cite{Radyushkin:1998rt,Diehl:1998kh} is to  use the concept of  
generalized parton distributions  (GPDs, see
  \cite{Goeke:2001tz,Diehl:2003ny,Belitsky:2005qn} for recent extensive reviews on GPDs) 
to describe/parametrize soft contributions.
The use of GPDs   allows to easily 
describe  existing data by soft contributions alone 
(the latest  attempts can be found in Refs.~\cite{Belitsky:2003nz,Diehl:2004cx,Guidal:2004nd}). 
 A subtle  point for  these semi-phenomenological 
approaches is to  avoid   double counting of    hard rescattering 
contributions  ``hidden'' in the model-dependent hadron wave functions
or GPD parametrizations.

The dominant role  of the soft contribution for the   pion form factor at  
moderate momentum transfers, up to $Q^2\sim 2-3$~GeV$^2$, is  supported by  its  
calculation \cite{Ioffe:1982qb,Nesterenko:1982gc}
within the QCD sum rule approach \cite{Shifman:1978bx}
applied to the  vacuum average $\langle 0 |T \{  \eta_2 (0) j(z) \eta_1^*(y) \} |0 \rangle$
 of  three currents, with $j$  representing the electromagnetic probe and the two others $ \eta_1^* , \eta_2$ 
having  quantum numbers of the initial and final hadrons, respectively.  
The application of the method at  higher $Q^2$ faces the 
problem  that the inclusion   of nonperturbative effects due to vacuum 
condensates
through  the expansion over  inverse powers of the Borel parameter $M^2$ 
interferes with the large-$Q^2$ expansion of the form factors, 
producing   a series of    $(Q^2/M^2)^n$ type even for a decreasing function 
of $Q^2$,  like $\exp[-Q^2/M^2]$.  
 For the nucleon form factors, the usual  QCD sum rule approach works only in the region 
 of small momentum transfers 
 $Q^2 <  1$~GeV$^2$ \cite{Belyaev:1992xf,Castillo:2003pt}. 
To extend the results  to higher $Q^2$, it was proposed \cite{Bakulev:1991ps} to sum  back 
the   $(Q^2/M^2)^n$ terms originating from  the  Taylor expansion 
of the same  nonlocal condensate, using to this end    simple models for it.  
 Another approach 
\cite{Nesterenko:1982gc,Nesterenko:1983ef}
is to use   the so-called local quark-hadron duality approximation,
in which the  values  of the  duality intervals are postulated to be 
$Q^2$-independent and are  taken from the two-point QCD sum rules.
The parameter-free results for the pion 
and nucleon form factors obtained in this way are in 
a rather good agreement with existing data.     

\begin{figure}[ht]
\centerline{\epsfxsize5cm\epsffile{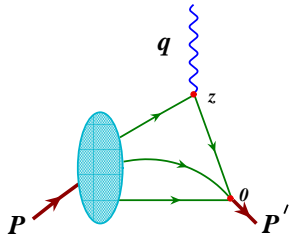}}
\caption{\label{figsum}\small
Schematic structure of the light-cone sum rule for baryon form factors.
}
\end{figure}

A less assumption-dependent approach allowing to calculate hadronic form factors
for moderately large $Q^2$ is based 
on light-cone  sum rules (LCSR) \cite{Balitsky:1989ry,Chernyak:1990ag}.  
The basic object of the LCSR approach is the 
matrix element $\langle 0| T \{ \eta_2(0) j(z) \} | h_1 \rangle $ in which the
 currents $\eta_2$ and $ j$ are the same as in the usual QCD sum rules,
 while the initial  hadron  is explicitly represented by its state vector 
 $| h_1 \rangle $, see a schematic representation in Fig.~\ref{figsum}.
 When both  the momentum transfer  $Q^2$ and 
 the Borel parameter $M^2$ of the $\eta_2$ channel are large,
 the asymptotics is governed by the operator product expansion   
$T \{ \eta_2(0) j(z) \} \sim \sum C_i(y) {\cal O}_i(0)$ on the 
light-cone $z^2=0$.   The  $z^2$-singularity  of a particular perturbatively calculable
short-distance factor  $C_i(z)$  is determined by the twist of the relevant
composite operator ${\cal O}_i$, whose matrix element $\langle 0|  {\cal O}_i(0)| h_1 \rangle $
accumulates  nonperturbative 
information about the initial state parametrized in terms of  a  
distribution amplitude. 
The lowest order $O(\alpha_s^0)$ terms of the OPE correspond to   
purely soft contributions ordered by twists of the relevant operators. The  magnitude 
and details of their $Q^2$ -dependence are governed by the 
form of the corresponding DA. As shown by the  LCSR 
calculation  of the pion form factor \cite{Braun:1994ij}, 
taking the  CZ model for the  pion DA, which enhances the 
hard contribution, one obtains a soft contribution 
that  is  too large.  Hence one should 
take the DAs that are sufficiently  narrower\footnote{Recent studies of the  pion DA 
\cite{Petrov:1998kg,Schmedding:1999ap,Anikin:2000rq,Bakulev:2001pa,Praszalowicz:2001wy,Bijnens:2002mg,Bakulev:2002uc,Bakulev:2004cu,Ball:2004ye,Agaev:2005gu,Agaev:2005rc}
 show  a converging consensus that 
the integral $\int_0^1 dx \, x^{-1} \varphi_{\pi}(x)$ determining the size 
of the leading-order hard contribution has the value close
to that corresponding to the asymptotic wave function,
see Ref.  \cite{Bakulev:2005cp} for an updated compilation.}  
in order to describe 
the  experimental  magnitude of the pion form factor 
at accessible $Q^2$.   
The LQSR expansion also  contains terms  
generating the asymptotic pQCD contributions. They   appear 
at proper order in $\alpha_s$, i.e., in  the $O(\alpha_s)$ term for the
pion form factor, at the $\alpha_s^2$ order for the nucleon form factors, etc. 
In the pion case, it was explicitly demonstrated 
\cite{Braun:1999uj,Bijnens:2002mg} that the contribution of hard  
rescattering is correctly reproduced in the LCSR   
approach as a part of the $O(\alpha_s)$ correction.
It should be noted that  the  diagrams of LCSR that 
contain the ``hard'' pQCD  contributions also possess ``soft'' parts,
i.e., one should perform  separation  of ``hard'' and ``soft''
terms inside each diagram.  As a result, 
the distinction between ``hard'' and ``soft'' contributions appears to 
be scale- and scheme-dependent \cite{Braun:1999uj}. 
During the  last years there have been numerous applications of LCSRs  
to mesons, see \cite{Braun:1997kw,Colangelo:2000dp} for a review.
 Nucleon electromagnetic form factors 
were first considered in \cite{Braun:2001tj,Lenz:2003tq} 
and the weak decay $\Lambda_b\to p\ell\nu_\ell$ in
\cite{Huang:2004vf}.  

In this paper, we  incorporate  light-cone  sum rules to  develop an 
approach to the calculation of transition form factors 
 for electroproduction of the $\Delta$-resonance. 
All three  possibilities for the virtual photon polarization
are allowed in this case, hence 
the $\gamma^* N \to \Delta$  transition is described by three 
independent form 
factors\footnote{The supply  of possible choices  and definitions of the three form factors 
offered  in  the literature seems inexhaustible.}. 
 Hence, the challenge is not only to fit 
the  absolute magnitude of one of them, but also to explain 
the relations between different form factors. 
In particular,  pQCD prediction 
\cite{Carlson:1985mm,Carlson:1988gt} for the 
helicity amplitudes is $A_{1/2}\sim 1/Q^3$ 
and $A_{3/2}\sim 1/Q^5$. For multipole amplitudes
$M1= -\frac12( A_{1/2}  + \sqrt{3} A_{3/2})$ and 
$E2= -\frac12(A_{1/2} -  A_{3/2}/\sqrt{3})$, 
the perturbative QCD approach   predicts, hence,  
the same strength of the  $E2$ and $M1$ 
transitions 
at asymptotically large $Q^2$. 
Experimentally,  their ratio is negative and extremely close to zero (within a few per cent) 
 in the whole investigated region, i.e.,  up to $Q^2 \sim $4\,GeV$^2$ 
\cite{Beck:1997ew,Frolov:1998pw,Kamalov:1999hs,Kamalov:2000en,Sato:1996gk}. 
To explain this phenomenon 
within the  pQCD framework, 
it was suggested \cite{Carlson:1998bg} that the observed small
value of the $E2/M1$ ratio  is due to cancellation of the 
$A_{1/2}$ and  $A_{3/2}$ helicity amplitudes.  
It should be emphasized that there is no intrinsic reason inside  perturbative QCD 
for such a cancellation to happen. 
Moreover, given the very strong sensitivity of the pQCD  outcome on the 
form of the nucleon and $\Delta$-isobar  distribution amplitudes 
(e.g., by varying 
the CZ type DAs without changing their moments within the 
limits specified by QCD sum rule estimates, 
one can change the result for  $A_{1/2}$ 
by  an order of magnitude and even reverse its  
sign\footnote{In this connection, we would also like to remark  that  
the  model $A_{1/2}(Q^2) = A_{1/2}(0)/(1+Q^2/\Lambda_1^2)^2$ 
proposed and used in Ref.~\cite{Carlson:1998bg}, with experimental
{\it negative} value of  $A_{1/2}(0)$, is not able 
to reproduce the  {\it positive} pQCD asymptotic results for 
$A_{1/2}(Q^2)$ quoted   in Eqs. (5)
and (6) of that paper.}, see Refs.~\cite{Carlson:1988gt,Carlson:1986zs,Stefanis:1992nw})
such a fine-tuned cancellation looks like a miracle.
Furthermore, since $A_{1/2}$  and  $A_{3/2}$ are expected to behave
differently with $Q^2$, just a  nearly constant   value 
of  $E2/M1$ ratio in a  sufficiently wide  region  of $Q^2$ (say, about $1$\,GeV$^2$ wide)  
 practically  rules out the relevance of pQCD within such a region.

On the other hand, the smallness of $E2/M1$ ratio is a famous prediction  of 
  the  quark model.
    40 years ago it was shown \cite{Becchi:1965}  that
 $E2$ is zero in the  nonrelativistic SU(6) quark model,
 provided that quarks have zero orbital angular momentum. 
  Small deviation of $E2$ from zero was later    explained either 
 by  $D$-wave admixtures \cite{Gershtein:1981zf,Isgur:1981yz}
 or in terms of two-body exchange currents \cite{Buchmann:1996bd}.
 In the large $N_c$ limit of QCD, it is possible to show \cite{Jenkins:2002rj}
 that the $E2/M1$ ratio has a smallness of order $1/N_c^2$ without making 
 any assumptions about the angular momentum of the quarks.
 Small values for   $E2/M1$ in the region $Q^2 < 4$\,GeV$^2$ were 
 obtained in the relativistic quark model \cite{Aznaurian:1993rk}.
  The $\gamma^*N \to \Delta$ transition form factors were also
  calculated \cite{Belyaev:1995ya}
  within the local quark-hadron duality approach motivated 
  by QCD sum rules.   It gives small values, within $\pm 20 \% 
  $, 
  both for the ratios $E2/M1$ and $C2/M1$ ($C2$ being the 
  electric quadrupole or Coulombic transition form factor). 
  Recent lattice calculations 
  \cite{Alexandrou:2003ea,Alexandrou:2004xn,Alexandrou:2005em}
  of the $N \Delta$ transition form factors up to 1.5\,GeV$^2$ 
  gives small negative values for both these ratios. 
  All these results provide strong evidence that the observed 
  small  value of $E2/M1$ has a purely nonperturbative origin.

Another interesting feature of the $\gamma^*N \to \Delta$ reaction 
is that the leading $M1$  magnetic transition 
form factor $G_{M} (Q^2)$ decreases with $Q^2$ faster
than the  dipole fit (see, e.g., \cite{Stoler:1993yk}). 
This was considered as a fact favoring pQCD since 
it usually gives for $G_{M} (Q^2)$ a  result  that 
is much smaller (by an order of magnitude) 
than that for the proton elastic form factor
$G_M^p (Q^2)$. 
In the large $N_c$ limit, assuming chiral and isospin 
symmetry,  it was established 
\cite{Frankfurt:1999xe,Goeke:2001tz}
that the transition form factor $G_M (Q^2)$
is expressed  through 
the isovector component of the 
GPD $E$  related to the elastic spin-flip nucleon 
form factor $F_2(Q^2)$, which is also known to
drop faster than the dipole fit. 
This observation was incorporated  in Ref.~\cite{Stoler:2002im}
to describe both $F_2(Q^2)$ and $G_{M} (Q^2)$
using a  
model for the GPDs  $E^{u,d} (x, Q^2)$
with a Gaussian $\sim \exp[-(1-x)Q^2/4x \lambda^2]$ 
plus a small power-law tail
ansatz for the $Q^2$-dependence. 
A Regge type ansatz  
$E^{u,d} (x, Q^2) =e^{u,d}(x) \, x^{\alpha'(0)(1-x)Q^2} $ 
 was used in Ref.~\cite{Guidal:2004nd}.
 In both models, to get an accurate fit of the data, one needs 
 to introduce a rescaling factor $\sim 1.5$ 
 in the relation between elastic and
 transition GPDs.

The goal of the present study is to set up the LCSR-based framework 
for the  calculation 
of  the $\gamma^*N \to \Delta$ transition form factors. 
The light-cone sum rule formalism turns out to be considerably more 
cumbersome in this case because of the spin 3/2 of the $\Delta$
resonance. The local interpolating current with 
the $\Delta$ quantum numbers has also a nonzero 
projection on the spin 1/2 states with 
opposite parity, and the necessity to get rid of these contaminating 
contributions  produces further complications. 
Still, as we show, it is possible to derive a general Lorentz 
decomposition and the twist expansion of  sum rules 
for all three form factors in question. 
We explicitly calculate 
the leading $\alpha_s$ order 
sum rules  and compare their consequences  with 
existing experimental data.   

Apart from resolving several technical issues, 
our main finding in this work  
is that the soft contribution to the  $\gamma^*N \to \Delta$  
form factors in the intermediate $Q^2$ range  
is strongly affected by the valence quark configurations 
involving ``minus'' components of the quark fields that do not 
have the simple interpretation in terms of the leading-twist amplitude but
rather correspond to the contributions of the orbital angular 
momentum\cite{Ji:2002xn,Ji:2003yj}.
The same conclusion was reached in \cite{Braun:2001tj}
for the nucleon electromagnetic form factor. In a more general context, 
large contributions of the orbital angular momentum can explain why  
helicity selection rules in perturbative QCD appear to be badly  
broken in hard exclusive processes at present energies.      
By construction, the LCSRs use nucleon 
distribution amplitudes of the leading and higher twist \cite{Braun:2000kw}
 as the main input, and the results are very sensitive to their shape.
Using the asymptotic distribution amplitudes we obtain a reasonable agreement
with the data in the range  $2<Q^2<6$~GeV$^2$ for the form factors.  
We believe that the accuracy can be improved significantly by the  
calculation of $O(\alpha_s)$ corrections to the sum rules and especially  
if lattice data on the moments of higher-twist distribution amplitudes  
become available.  
A long-term goal of our study is to determine 
leading twist nucleon distribution amplitudes
from the combined fit to the experimental 
data on all of the existing form factors involving the nucleon.
In this  perspective, 
the present paper should be viewed as a step in this direction.
  
The presentation  is organized as follows.
In Section 2  we introduce the necessary notation and set up the 
general framework. 
Section 3 contains the derivation of sum rules including higher 
twist corrections, which is our main result.
The numerical analysis of the LCSRs is carried out in Section~4,
together with a summary and discussion.
The paper has three Appendices devoted to technical aspects of the 
calculation: In Appendix A we present a complete Lorentz-invariant 
decomposition of the correlation function, Appendix B contains
the summary of asymptotic expressions for the nucleon distribution amplitudes,
and  in Appendix C  the Belyaev-Ioffe 
sum rule for the $\Delta$ coupling constant is given and reanalyzed.

%%%%%%%%%%%%%%%%%%%%%%%%%%%%
\setcounter{equation}{0} \section{General Framework}
%%%%%%%%%%%%%%%%%%%%%%%%%%%%

\subsection{Definition of the form factors}

The  $\gamma^* N \to \Delta$ transition
is described by the  matrix element of the electromagnetic current 
\beq{EM}
j_{\nu}=e_{u}\overline{u}\gamma_{\nu}u+e_{d}\overline{d}\gamma_{\nu}d
\eeq
between the nucleon state with momentum $P$ and the $\Delta$-isobar 
state with momentum $P'=P-q$. It can be  written as 
\beq{Ndelta}
\langle \Delta(P') \mid j_{\nu}(0)\mid N(P)\rangle = 
\bar{\Delta}_{\beta}(P')\Gamma_{\beta\nu}\gamma_5 N(P)\, ,
\eeq
where $N(P)$ and $\Delta_\beta(P')$ are the nucleon spinor and the 
Rarita-Schwinger spinor for the $\Delta$-isobar, 
respectively\footnote{We hope that denoting particle states by 
the same letters as the corresponding spinors will not 
create confusion.}. 
The decomposition  of the vertex function 
\bea{G123}
\Gamma_{\beta\nu} &=& G_{1}(Q^{2})\big[-q_{\beta}\gamma_{\nu}+q\s g_{\beta\nu}\big]
+G_{2}(Q^{2})\big[-q_{\beta}(P-q/2)_{\nu}+q(P-q/2)g_{\beta\nu}\big]
\nonumber\\&&{}
+G_{3}(Q^{2})\big[q_{\beta}q_{\nu}-q^{2}g_{\beta\nu}\big].
\eea
 defines   three scalar form factors $G_i  (Q^2)$. As usual, $Q^2 =-q^2$. 
Following \cite{Jones:1972ky},  one can also define the magnetic dipole $G_{M}$, 
electric quadrupole $G_{E}$, and Coulomb quadrupole $G_{C}$ form 
factors instead of $G_{1}$, $G_{2}$, $G_{3}$:
\bea{MEQ}
G_{M}(Q^{2})&=&
\frac{m_{P}}{3(m_{P}+m_{\Delta})}\Bigg[((3m_{\Delta}+m_{P})(m_{\Delta}+m_{P})+Q^{2})\frac{G_{1}(Q^{2})}{m_{\Delta}}
\nonumber\\
&&\hspace{2cm}+(m_{\Delta}^{2}-m_{P}^{2})G_{2}(Q^{2})-2Q^{2}G_{3}(Q^{2})\Bigg],
\nonumber\\
G_{E}(Q^{2})&=&
\frac{m_{P}}{3(m_{P}+m_{\Delta})}\Bigg[(m_{\Delta}^{2}-m_{P}^{2}-Q^{2})\frac{G_{1}(Q^{2})}{m_{\Delta}}
\nonumber\\
&&\hspace{2cm}+(m_{\Delta}^{2}-m_{P}^{2})G_{2}(Q^{2})-2Q^{2}G_{3}(Q^{2})\Bigg],
\nonumber\\
G_{C}(Q^{2})&=&
\frac{2m_{P}}{3(m_{\Delta}+m_{P})}\Bigg[2m_{\Delta}G_{1}(Q^{2})
  +\frac{1}{2}(3m_{\Delta}^{2}+m_{P}^{2}+Q^{2})G_{2}(Q^{2})
\nonumber\\
&&\hspace{2cm}+(m_{\Delta}^{2}-m_{P}^{2}-Q^{2})G_{3}(Q^{2})\Bigg] \,  .
\eea
For a comparison with the literature we also write down 
the form factors $G_M^{\rm Ash}$ \cite{Ash}, $G_T$ \cite{Stoler:1993yk} and the ratios
$R_{EM}$ \cite{Jones:1972ky} and $R_{SM}$ (see e.g. \cite{Buchmann:2004ia,Caia:2004pm,Pascalutsa:2005ts})
that are used in experimental papers
\bea{GT}
 G_M (Q^{2})&=& G_M^{\rm Ash}(Q^{2})\sqrt{1+\frac{Q^2}{(m_\Delta+m_P)^2}}\,,
\nonumber\\
\left|G_{M}(Q^{2})\right|^{2}+3\left|G_{E}(Q^{2})\right|^{2}&=&
\frac{Q^{2}}{Q^{2}+\nu^{2}}\Bigg(1+\frac{Q^{2}}{(m_{\Delta}+m_{P})^{2}}\Bigg)\left|G_{T}(Q^{2})\right|^{2},
   \quad \nu =  \frac{m_{\Delta}^{2}-m_{P}^{2}+Q^{2}}{2m_{P}},
\nonumber \\
R_{EM}(Q^2) & = & \frac{E2(Q^{2})}{M1(Q^{2})} = \frac{E_{1+}}{M_{1+}} = - \frac{G_E(Q^2)}{G_M(Q^2)}
 \\
R_{SM}(Q^2) & = & \frac{C2(Q^{2})}{M1(Q^{2})} = \frac{S_{1+}}{M_{1+}} 
= -\sqrt{Q^2 +\frac{(m_{\Delta}^2- m_p^2-Q^2)^2}{4 m_{\Delta}^2}}
\frac{1}{2 m_{\Delta}} \frac{G_C(Q^2)}{G_M(Q^2)}
\, .
\nonumber
\eea
Note that there is a disagreement in the literature on the overall sign in the relation between $R_{SM}$ and the ratio
of the quadrupole and the magnetic form factors $G_C(Q^2)/G_M(Q^2)$. We follow the definition from  \cite{Pascalutsa:2005ts}. 

%%%%%%%%%%%%%%%
\subsection{Choice of the correlation function}
%%%%%%%%%%%%%%

To extract information about  hadronic form  factors within the light-cone sum rule 
approach  we should analyze a matrix element in which  one of the hadrons is represented 
by an interpolating field  with   proper quantum numbers, and another  is described 
by its explicit state vector, cf. Fig.~\ref{figsum}. 
 Building the sum rule, we  would need distribution amplitudes of the latter  hadron. 
The  nucleon  DAs for all three-quark operators were 
introduced and studied  in Ref.~\cite{Braun:2000kw}, while 
in the case of the $\Delta$-isobar such an analysis is yet to be performed.
 Thus, in the present  paper, we   consider the 
correlation function given by the matrix element 
\beq{T}
T_{\mu\nu}(P,q)=i\int d^{4}z\, e^{iqz} \langle0 
| T\left\{\eta_{\mu}(0)j_{\nu}(z)\right\} | N(P)\rangle
\eeq
between the vacuum and a single-nucleon state $| N(P)\rangle$.   
The interpolating field for 
the $\Delta^{+}$-particle is taken in the form suggested in \cite{Ioffe:1981kw}
\beq{eta}
\eta_{\mu}(0)= 
 \epsilon^{abc}\left[2(u^{a}(0)C\gamma_{\mu}d^{b}(0))u^{c}(0)+
 (u^{a}(0)C\gamma_{\mu}u^{b}(0))d^{c}(0)\right]\, , 
\eeq
where $a,b,c$ are color indices and $C$ is the charge conjugation matrix. 
The contribution of $\Delta^+$ to the correlation function in \Gl{T} is given by  
\beq{Tdelta}
T_{\mu\nu}(P,q)=
\frac{1}{m_{\Delta}^{2}-(P')^{2}}\sum_{s}\langle 0 | \eta_{\mu}(0)| \Delta(P',s)\rangle\langle 
\Delta(P',s) | j_{\nu}(0)| N(P)\rangle. 
\eeq
Parametrizing  the matrix element 
\beq{lambdaD}
\langle0|\eta_{\mu}(0)| \Delta(P',s)\rangle = \frac{\lambda_{\Delta}}{(2\pi)^{2}} \Delta^{(s)}_{\mu}(P')
\eeq
in terms of   $\lambda_{\Delta}$,   
the coupling  constant of the $\Delta^{+}$-particle to the current $\eta_\mu$, 
and using the standard spin summation formula for Rarita-Schwinger spinors
\beq{spinsum}
\sum_{s}\Delta_{\mu}^{(s)}(P')\overline{\Delta}_{\beta}^{(s)}(P')
   = -(\pslash'+m_{\Delta})\left\{g_{\mu\beta}-\frac13\gamma_{\mu}\gamma_{\beta}
     -\frac{2P'_{\mu}P'_{\beta}}{3m_{\Delta}^{2}}+\frac{P'_{\mu}\gamma_{\beta}
     -P'_{\beta}\gamma_{\mu}}{3m_{\Delta}}\right\} \, 
\eeq
we write this contribution  as 
\bea{T32}
\lefteqn{T_{\mu\nu}^{(\Delta)}(P,q)=}
\nonumber\\&=&{}
        -\frac{\lambda_{\Delta}}{(2\pi)^{2}}\frac{\pslash'+m_{\Delta}}{m_{\Delta}^{2}-(P')^{2}}
   \left[g_{\mu\beta}-\frac{1}{3}\gamma_{\mu}\gamma_{\beta}
 -\frac{2P'_{\mu}P'_{\beta}}{3m_{\Delta}^{2}}+ 
\frac{P'_{\mu}\gamma_{\beta}-P'_{\beta}\gamma_{\mu}}{3m_{\Delta}}\right] \Gamma_{\beta\nu}\gamma_5 N(P) \,  . 
\eea
 This expression provides us with  one of  the starting points of our analysis. 

To extract the $\Delta$-related information from
the total correlator $T_{\mu\nu}(P,q)$, we should take into account 
a subtlety  that the current $\eta_\mu$ couples 
not only to isospin $I=\frac32$ spin-parity $\frac32^+$ states, 
but also to isospin $I=\frac32$ spin-parity $\frac12^-$ states.  
It makes sense  to eliminate their contribution
by a clever choice of the Lorentz structures. 
For a generic $\frac12^-$ resonance, call it $\Delta^*$, 
we define
\beq{12}
 \langle0|\eta_{\mu}(0)|\Delta^*(P',s)\rangle=\frac{\lambda_{*}}{(2\pi)^2}
 (m_{*}\gamma_{\mu}-4P'_{\mu})\Delta^{*}(P',s)
\eeq
with $\lambda^{*}$ being  the corresponding coupling and $m_{*}$ the mass.  The $\Delta^*$ spinor satisfies 
the usual Dirac equation $(\not\!\!P'-m_{*})\Delta^{*}(P')=0$, and the matrix element of the electromagnetic 
current between the nucleon and a $\Delta^*$ state takes the form 
\beq{EM12}
 \langle \Delta^*(P')| j_{\nu}(0) | N(P)\rangle = \overline{\Delta^*}(P')
\left[
   (\gamma_{\nu}q^{2}-\not\!q q_{\nu})F^{N\Delta^*}_1(q^2) -i\sigma_{\mu\alpha}q^{\alpha}F^{N\Delta^*}_{2}(q^2)
      \right]\gamma_5 N(P)\,. 
\eeq
The corresponding contribution to the correlation function in \Gl{T} reads 
\bea{T12}
\lefteqn{T_{\mu\nu}^{({\Delta^*})}(P,q)=}
\nonumber\\&=& 
\frac{\lambda^*_{\Delta}}{(2\pi)^{2}}\big[m_{*}\gamma_{\mu}-4P'_{\mu}\big]
\frac{\not \! P^{\prime}+m_{*}}{m_{*}^{2}-(P')^{2}}
\left[(\gamma_{\nu}q^{2}-\not\!q q_{\nu})F^{N\Delta}_{1}-
 i\sigma_{\nu\alpha}q^{\alpha}F^{N\Delta}_{2}\right]\gamma_{5}N(P) \,  ,
\eea
and the question is whether (and at which cost) the contributions in \Gl{T32} and \Gl{T12} can be separated.
To achieve this,  we  have to understand the general decomposition of the correlation function in \Gl{T}
into contributions of different Lorentz structures.     
To this end, it is convenient to go over to the twist decomposition in the infinite momentum frame.

%%%%%%%%%%%%%%%%%
\subsection{Light-Cone Expansion}
%%%%%%%%%%%%%%%%%

\subsubsection{Kinematics}

Having in mind the practical construction of light-cone sum rules that involve nucleon distribution amplitudes,
we define a light-like vector $n_\mu$ by the condition
\beq{z}
       q\cdot n =0\,,\qquad n^2 =0
\eeq
and introduce the second light-like vector 
vector 
\bea{vectors}
p_\mu &=& P_\mu  - \frac{1}{2} \, n_\mu \frac{m_P^2}{P\cdot n}\,,~~~~~ p^2=0\,, 
\eea
so that $P \to p$ if the nucleon mass can be neglected, $m_P \to 0$.
The photon momentum  then can be written as 
\begin{eqnarray}
q_{\mu}=q_{\bot \mu}+ n_{\mu}\frac{P\cdot q}{P\cdot n}\, .
\end{eqnarray}
Assume for a moment that the nucleon moves in the positive 
${\bf e_z}$ direction, then $p^+$ and $n^-$ are the only nonvanishing 
components of $p$ and $n$, respectively. 
The infinite momentum frame can be visualized
as the limit $p^+ \sim Q \to \infty$ with fixed $P\cdot n = p \cdot n\sim 1$ 
where $Q$ is the large scale in the process.
Expanding the matrix element in powers of $1/p^+$ introduces
the power counting in $Q$. In this language,  twist counts the
suppression in powers of $p^+$. Similarly, 
the nucleon spinor $N_\ga(P,\la)$ has to be decomposed 
in ``large'' and ``small'' components as
\beq{spinor}
N_\ga(P,\la) = \frac{1}{2 p\cdot n} \left(\!\not\!{p}\! \!\not\!{n} +
\!\not\!{n}\!\!\not\!{p} \right) N_\ga(P,\la)= N^+_\ga(P,\la) + N^-_\ga(P,\la)
\; ,
\eeq
where we have introduced two projection operators
\beq{project}
\Lambda^+ = \frac{\!\not\!{p}\! \!\not\!{n}}{2 p\cdot n} \quad ,\quad
\Lambda^- = \frac{\!\not\!{n}\! \!\not\!{p}}{2 p\cdot n}
\eeq
that project onto the ``plus'' and ``minus'' components of the spinor.
Note  the useful relations
\beq{bwgl}
\lash{p} N(P) = m_P N^+(P)\,,\qquad \lash{n} N(P) = \frac{2 p \cdot n}{m_P} N^-(P)
\eeq
that follow readily from the Dirac equation $\not\!\!{P} N(P) = M N(P)$.
Using the explicit expressions for $N(P)$ it is easy to see 
that $\Lambda^+N = N^+ \sim \sqrt{p^+}$ while $\Lambda^-N = N^- \sim 1/\sqrt{p^+}$.

The correlation function $T_{\mu\nu}$ in \Gl{T} can be expanded in contributions with increasing twist.
The leading twist contribution corresponds to the projection 
\beq{lt}
        \Lambda_+ T_{++}
\eeq
and contains the maximum  power of the large momentum $p^+$. There exist three projections
of the next-to-leading twist that are suppressed by one power of 
the large momentum compared to the leading twist, namely 
\bea{t+1}
&& \Lambda_- T_{++}\,,\quad \Lambda_+ T_{\perp +}\,,\quad \Lambda_+ T_{+\perp} \, ,
\eea
and more contributions with loosing two powers of the momentum, etc.

Each light-cone projection in a general situation gives rise to several invariant 
Lorentz structures, in particular
\bea{decomp}
%  \frac{p\s n\s}{2pn}
\Lambda_+\,n^{\mu}n^{\nu}T_{\mu\nu}&=& (p \cdot n)^2 
   \Big\{ {\cal A}(Q^2,Pq) + \not\!q_{\bot}\,{\cal B}(Q^2,Pq) \Big\}\gamma_{5}N^{+}(P)\,,
\nonumber\\
%    \frac{n\s p\s}{2pn}
\Lambda_-\,n^{\mu}n^{\nu}T_{\mu\nu}&=& (p \cdot n)^2 
   \Big\{ {\cal C}(Q^2,Pq) + \not\!q_{\bot}\,{\cal D}(Q^2,Pq) \Big\}\gamma_{5}N^{-}(P)\,,
\nonumber\\
%\frac{p\s n\s}{2pn}
\Lambda_+\,n^{\mu}e_{\bot}^{\nu}T_{\mu\nu}&=& 
 (p \cdot n)\Big\{ 
     \not\!e_{\bot}\,{\cal E}(Q^2,Pq) +  \not\!e_{\bot}\not\!q_{\bot}\,{\cal F}(Q^2,Pq)\Big\}\gamma_{5}N^{+}(P)\,,
\nonumber\\
\Lambda_+\,e_{\bot}^{\mu}n^{\nu}T_{\mu\nu}&=& 
 (p \cdot n)\Big\{
    \not\! e_{\bot}\,{\cal G}(Q^2,Pq) +  \not\!e_{\bot}\not\!q_{\bot}\,{\cal H}(Q^2,Pq)\Big\}\gamma_{5}N^{+}(P)\,,
\nonumber\\
\Lambda_+\,q_{\bot}^{\mu}n^{\nu}T_{\mu\nu}&=& 
 (p \cdot n)\Big\{{\cal I}(Q^2,Pq) + \not\!q_{\bot}\,{\cal J}(Q^2,Pq) \Big\}\gamma_{5}N^{+}(P)\,,
\eea
where $e_{\bot}$ is a two-component vector in the transverse plane that is 
orthogonal to $q_{\bot}$:
\beq{ee}
  e_{\bot}\cdot q_{\bot} =0\,,\quad (e_{\bot})^2 = 1\,.   
\eeq
The invariant amplitudes ${\cal A}$\,--\,${\cal J}$ are not all independent. First of all, note 
that the $n^\mu q_{\bot}^{\nu}T_{\mu\nu}$-projection does not lead to new independent amplitudes
because of the transversality condition $q^\nu T_{\mu\nu}=0$. Second, the 
Rarita-Schwinger constraint $\gamma^\mu T_{\mu\nu}=0$ results in   two relations
\bea{RScond}
   {\cal G} + {\cal J} - m_P\, {\cal C} &=&0\,,
\nonumber\\
  {\cal H}- \frac{1}{Q^2}{\cal I} + m_P\, {\cal D} &=&0\,.
\eea
Finally, more relations follow from the Lorentz symmetry. In order to find them,  we write 
a Lorentz-invariant decomposition of the correlation function $T_{\mu\nu}$ which involves 
10 independent amplitudes, see Appendix~A. Taking the necessary light-cone projections,
 we  obtain two new relations that can be chosen as 
\bea{LorSym}
  2\, {\cal F} -m_P\, {\cal D} + m_P\, {\cal B} + 2\, {\cal H} &=& 0\,.
\nonumber\\
 m_P\,{\cal A} -2(Pq){\cal B}- m_P\,{\cal C} -2\, {\cal E}-2\, {\cal G} &=& 0\,.
\eea
The remaining 6 independent invariant functions produce an overcomplete 
set of sum rules of the leading and next-to-leading twist 
for the three  form factors of the $\gamma^* N \to \Delta$ transition.

\subsubsection{The $J^P=\frac32^+$ contributions}

Now, the contribution of the $\Delta^+$ to the   
invariant functions defined above can readily be calculated 
using Eq.~(\ref{T32}). 
The result has the following form: 
\bea{lhs}
   {\cal A}^{\Delta}&=& 
-\frac{\lambda_{\Delta}/(2\pi)^2}{(m_{\Delta}^{2}-P'^{2})} \frac{Q^2}{3m_{\Delta}^{2}}
\Big[ 2G_{1}+(2m_{\Delta}-m_{P})G_{2} +2(m_{P}-m_{\Delta})G_{3}\Big],
\nonumber\\
      {\cal B}^{\Delta} &=& 
-\frac{\lambda_{\Delta}/(2\pi)^2}{(m_{\Delta}^{2}-P'^{2})} \frac{1}{3m_{\Delta}^{2}}
\Big[-2m_{P}G_{1}+[m_{\Delta}(m_{\Delta}-m_{P})-Q^{2}]G_{2}+2Q^{2}G_{3}\Big],
\nonumber\\
   {\cal C}^{\Delta}&=& 
\frac{\lambda_{\Delta}/(2\pi)^2}{(m_{\Delta}^{2}-P'^{2})} \frac{Q^2}{3m_{\Delta}^{2}m_P}
\Big[2(m_{P}+m_{\Delta})G_{1}+(m_{\Delta}^{2}+Q^{2})G_{2}+2[m_{\Delta}(m_{P}-m_{\Delta})-Q^{2}]G_{3}\Big],
 \nonumber\\
{\cal D}^{\Delta}&=&-\frac{\lambda_{\Delta}/(2\pi)^2}{(m_{\Delta}^{2}-P'^{2})}
\frac{1}{3m_{\Delta}^{2}m_{P}}
\Big[2(Q^{2}
-m_{\Delta}m_{P})G_{1}
+(m_{\Delta}-m_{P})(m_{\Delta}^{2}+Q^{2})G_{2}
\nonumber\\
&&{}+2m_{P}Q^{2}G_{3}\Big],
\nonumber\\
{\cal E}^{\Delta}&=&-\frac{\lambda_{\Delta}/(2\pi)^2}{(m_{\Delta}^{2}-P'^{2})}
\frac{1}{6m_{\Delta}^{2}}
 \Big[2[m_{P}(m_{P}^{2}-m_{\Delta}^{2})+Q^{2}(2m_{\Delta}+m_{P})]G_{1}
\nonumber\\
&&{}+ m_{\Delta} (m_{\Delta}-m_{P})^2 (m_{\Delta}+m_{P}) G_{2}
+2Q^{2}m_{\Delta}(m_{P}-m_{\Delta})G_{3}\Big],
\nonumber\\
{\cal F}^{\Delta}&=&-\frac{\lambda_{\Delta}/(2\pi)^2}{(m_{\Delta}^{2}-P'^{2})}
\frac{1}{6m_{\Delta}^{2}}
\Big[2[Q^{2} -4m_{\Delta}^{2}+ (m_{\Delta}-m_{p})^2]G_{1}
+m_{\Delta}(m_{P}^{2}-m_{\Delta}^{2})G_{2}
\nonumber\\
&&{}+2m_{\Delta}Q^{2}G_{3}\Big],
\nonumber\\
{\cal G}^{\Delta}&=&
-\frac{\lambda_{\Delta}/(2\pi)^2}{(m_{\Delta}^{2}-P'^{2})}
\frac{Q^2}{6m_{\Delta}}
\Big[-2G_{1}+ (m_{P}+m_{\Delta})G_{2}+2(m_{\Delta}-m_{P})G_{3}\Big],
\nonumber\\
{\cal H}^{\Delta}&=&-\frac{\lambda_{\Delta}/(2\pi)^2}{(m_{\Delta}^{2}-P'^{2})}
\frac{1}{6m_{\Delta}}
\Big[2(3m_{\Delta}+m_{P})G_{1}+(2m_{\Delta} (m_{\Delta}-m_{P}) +Q^{2})G_{2}-2Q^{2}G_{3}\Big],
\nonumber\\
{\cal I}^{\Delta}&=&
\frac{\lambda_{\Delta}/(2\pi)^2}{(m_{\Delta}^{2}-P'^{2})}
\frac{Q^2}{6m_{\Delta}^{2}}
\Big[2(m_\Delta m_P -3 m_{\Delta}^2 -2 Q^2)]G_{1}
\nonumber\\
&&{}+ (4m_\Delta^2 (m_P-m_\Delta) + Q^2(2 m_P-3m_\Delta))G_2+2Q^2(m_\Delta-2m_P)G_{3}\Big],
\nonumber\\
{\cal J}^{\Delta}&=&
-\frac{\lambda_{\Delta}/(2\pi)^2}{(m_{\Delta}^{2}-P'^{2})}
\frac{Q^2}{6m_\Delta^2}
\Big[-2(2m_{P} +m_{\Delta})G_{1}
-(m_\Delta (m_P +3 m_\Delta) + 2 Q^2)G_{2}
\nonumber\\
&&\hspace{1cm}+2(m_\Delta (m_\Delta- m_P) +2 Q^2)G_{3}\Big].
\eea
%%%%%%%%%%%%%%%%%%%%%%%%%%%%%%%%%%%%%%%%%%%%%%%%%%%%%%%%%%%%

\subsubsection{The $J^P=\frac12^-$ contributions}

{}For completeness, we present here the (unwanted) 
contributions to the same invariant functions ${\cal A}$\ldots ${\cal J}$
of the negative parity spin-1/2 $\Delta$-resonances,
cf.~(\ref{T12}):
\bea{lhs12}
{\cal A}^{\Delta^*}&=&\frac{8\lambda_{*}/(2\pi)^2}{(m_{*}^{2}-P'^{2})}\, Q^2 F_{1}\,,
\nonumber\\
{\cal B}^{\Delta^*}&=&-\frac{8\lambda_\ast/(2\pi)^2}{(m_{*}^{2}-P'^{2})}F_{2}\,,
\nonumber\\
{\cal C}^{\Delta^*}&=&-\frac{4\lambda_\ast/(2\pi)^2Q^{2}}{(m_{*}^{2}-P'^{2})m_{P}}\Big[m_* F_{1}+2 F_{2}\Big],
\nonumber\\
{\cal D}^{\Delta^*}&=&\frac{4\lambda_\ast/(2\pi)^2}{(m_{*}^{2}-P'^{2})m_{P}}\Big[2 Q^2 F_{1}-m_{*}F_{2}\Big],
\nonumber\\
{\cal E}^{\Delta^*}&=&\frac{4\lambda_\ast/(2\pi)^2}{(m_{*}^{2}-P'^{2})}
\Big[Q^{2}(m_{*}+m_{P})F_1+ (m_{P}^{2}-m_{*}^{2})F_2\Big],
\nonumber\\
{\cal F}^{\Delta^*}&=&\frac{4\lambda_\ast/(2\pi)^2}{(m_{*}^{2}-P'^{2})}\Big[Q^{2}F_1+(m_{P}-m_{*})F_2\Big],
\nonumber\\
{\cal G}^{\Delta^*}&=&-\frac{2\lambda_\ast/(2\pi)^2}{(m_{*}^{2}-P'^{2})}Q^2 \, m_* F_{1}\,,
\nonumber\\
{\cal H}^{\Delta^*}&=&\frac{2\lambda_\ast/(2\pi)^2}{(m_{*}^{2}-P'^{2})}\, m_* F_{2}\,,
\nonumber\\
{\cal I}^{\Delta^*}&=&\frac{2\lambda_\ast/(2\pi)^2}{(m_{*}^{2}-P'^{2})}Q^2 \, \Big[4Q^2F_1-m_*F_2\Big]\,,
\nonumber\\
{\cal J}^{\Delta^*}&=&-\frac{2\lambda_\ast/(2\pi)^2}{(m_{*}^{2}-P'^{2})}Q^2 \,\Big[m_*F_1+4F_2\Big]\,.
\eea
Here $F_{1,2}=F_{1,2}^{N\Delta^*}(Q^2)$ and $m_* = m_{\Delta^*}$ is the mass of the $\Delta^*$ resonance.
Taking into account the relations in (\ref{RScond}) and (\ref{LorSym})  
we find that there exist two independent combinations of the invariant functions that are 
free from contributions of $J^P=\frac12^-$ resonances  {\it with arbitrary mass $m_*$}:
\bea{free12}
    && {\cal G}^{\Delta^*} - {\cal J}^{\Delta^*} + Q^2 {\cal B}^{\Delta^*} =0\,,
\nonumber\\
    && {\cal I}^{\Delta^*} + Q^2\,{\cal H}^{\Delta^*} - Q^2 {\cal A}^{\Delta^*} =0\, .
\eea  
On the other hand, using (\ref{lhs}) one obtains very simple expressions for the $\Delta$-isobar 
contributions to the same combinations: 
\bea{free32}
   {\cal G}^{\Delta} - {\cal J}^{\Delta} + Q^2 {\cal B}^{\Delta} &=& 
- \frac{\lambda_\Delta/(2\pi)^2}{(m_\Delta^2 - P'^2)} Q^2 G_2(Q^2)\,,
\nonumber\\     
   {\cal I}^{\Delta} + Q^2\,{\cal H}^{\Delta} -  Q^2\,{\cal A}^{\Delta} &=&
- \frac{\lambda_\Delta/(2\pi)^2}{(m_\Delta^2 - P'^2)}Q^2\big[2 G_1(Q^2)+(m_\Delta-m_P)G_2(Q^2)\big].
\eea
This implies that one can construct light-cone sum rules for the $\gamma^* N \to \Delta$ form factors 
$G_1$ and $G_2$ that are free from contamination by negative parity spin-1/2 resonances. 
{}For the form factor $G_3$ this separation is not possible, unless one 
goes over to Lorentz structures of yet higher twist. Since the accuracy of light-cone sum 
rules is expected to decrease with twist, the gain of excluding spin-1/2 resonances in this
case is probably not worth the effort. Note that the difference between the magnetic $G_M$ and 
electric $G_E$ form factors only involves $G_1$:
\bea{GM-GE}
      G_M(Q^2)-G_E(Q^2) &=& \frac{2m_P}{3m_\Delta}\frac{[(m_\Delta+m_P)^2+Q^2]}{(m_\Delta+m_P)}
      G_1(Q^2)
\eea

%%%%%%%%%%%%%%%%%
\setcounter{equation}{0} \section{Calculation of  Correlation Functions}
%%%%%%%%%%%%%%%%%

On the other side, one can calculate the invariant functions ${\cal A}$\ldots ${\cal J}$ for Euclidean 
virtuality  $(P')^2 = (P-q)^2$ in terms of the nucleon (proton) DAs of increasing twist.
The calculation can be simplified considerably by observing that only the isospin-one 
part of the electromagnetic current can initiate $N\Delta$-transitions. Hence,  all correlation
functions must be proportional to $e_u-e_d$ and a (simpler) calculation of the $d$-quark 
contribution suffices to obtain the complete result. An additional simplification is due to the 
fact that, to the leading-order accuracy in QCD coupling, the two $u$-quarks in the $\Delta$-current 
(\ref{eta}) remain spectators and retain their position at the origin. It is, therefore,  sufficient
to retain only those terms in the general Lorentz decomposition of the three-quark matrix element
between vacuum and the proton state, that are symmetric with respect to the interchange of the momentum 
fractions of the two $u$-quarks \cite{Braun:2000kw}:
\bea{zerl}  
\lefteqn{ 4 \bra{0} \ep^{ijk} u_\al^i(0) u_\be^j(0) d_\ga^k(z)  
          \ket{N(P)} =}  
\nonumber\\  
&=&  
\left({\cal V}_1 + \frac{z^2m_P^2}{4} {\cal V}_1^M \right) \left(\!\not\!{P}C \right)_{\al \be} \left(\ga_5 N\right)_\ga +  
{\cal V}_2 m_P \left(\!\not\!{P} C \right)_{\al \be} \left(\!\not\!{z} \ga_5 N\right)_\ga  +  
{\cal V}_3 m_P  \left(\ga_\mu C \right)_{\al \be}\left(\ga^{\mu} \ga_5 N\right)_\ga  
\nn \\  
&&{} +  
{\cal V}_4 m_P^2 \left(\!\not\!{z}C \right)_{\al \be} \left(\ga_5 N\right)_\ga +  
{\cal V}_5 m_P^2 \left(\ga_\mu C \right)_{\al \be} \left(i \si^{\mu\nu} z_\nu \ga_5  
N\right)_\ga  
+ {\cal V}_6 m_P^3 \left(\!\not\!{z} C \right)_{\al \be} \left(\!\not\!{z} \ga_5 N\right  
)_\ga\,
\nonumber\\
&&{}+
\left({\cal T}_1+\frac{z^2m_P^2}{4}{\cal T}_1^M\right)
\left(P^\nu i \si_{\mu\nu} C\right)_{\al \be} \left(\ga^\mu\ga_5 N\right)_\ga 
+ 
{\cal T}_2 m_P \left(z^\mu P^\nu i \si_{\mu\nu} C\right)_{\al \be} \left(\ga_5 N\right)_\ga 
\nn \\
&&{} 
+ {\cal T}_3 m_P \left(\si_{\mu\nu} C\right)_{\al \be} \left(\si^{\mu\nu}\ga_5 N\right)_\ga 
+ {\cal T}_4 m_P \left(P^\nu \si_{\mu\nu} C\right)_{\al \be} \left(\si^{\mu\ro} z_\ro \ga_5 N\right)_\ga 
\nn \\
&&{} 
+ {\cal T}_5 m_P^2 \left(z^\nu i \si_{\mu\nu} C\right)_{\al \be} \left(\ga^\mu\ga_5 N\right)_\ga 
+{\cal T}_6 m_P^2 \left(z^\mu P^\nu i \si_{\mu\nu} C\right)_{\al \be} \left(\!\not\!{z} \ga_5 N\right)_\ga  
\nn \\&&
+{\cal T}_{7} m_P^2 \left(\si_{\mu\nu} C\right)_{\al \be} \left(\si^{\mu\nu} \!\not\!{z} \ga_5 N\right)_\ga
+ {\cal T}_{8} m_P^3 \left(z^\nu \si_{\mu\nu} C\right)_{\al \be} \left(\si^{\mu\ro} z_\ro \ga_5 N\right)_\ga\,.
\eea   
The expansion in (\ref{zerl}) should be viewed as an operator product expansion 
to the leading order in the strong coupling. 
Each of the 
functions ${\cal V}_i$ and  ${\cal T}_i$  
is a function of the scalar product $(P\cdot z)$ and 
also depends on the deviation from the light-cone $z^2$ at most 
logarithmically. In addition,  we  take into account the $O(z^2)$ corrections to the 
leading-twist-3 structures, denoted by ${\cal V}^M_1$ and ${\cal T}^M_1$. 
The invariant functions ${\cal V}_1(Pz),\ldots,{\cal T}_8(Pz)$ can be expressed in terms of the 
nucleon distribution 
amplitudes  $V_1(x_i),\ldots,T_8(x_i)$ with increasing twist, introduced in Ref.~\cite{Braun:2000kw}:
\bea{opev}  
\renewcommand{\arraystretch}{1.7}  
\begin{array}{lll}  
 {\cal V}_1 = V_1\,, &~~& 2 \,(pz) {\cal V}_2 = V_1 - V_2 - V_3\,, \\  
 2\, {\cal V}_3 = V_3\,, &~~& 4 \,(pz) {\cal V}_4 = - 2 V_1 + V_3 + V_4  + 2 V_5\,, \\  
4\, (pz)\, {\cal V}_5 = V_4 - V_3\,, &~~&  
4 \,(p z )^2  {\cal V}_6 = - V_1 + V_2 +  V_3 +  V_4 + V_5 - V_6\, \\
{\cal T}_1 = T_1\,, && 2 \, {\cal T}_2 = T_1 + T_2 - 2 T_3\,, 
 \\
2 {\cal T}_3 = T_7\,, && 2 \,(pz) {\cal T}_4 = T_1 - T_2 - 2  T_7\,, 
 \\
2 \,(pz) {\cal T}_5 = - T_1 + T_5 + 2  T_8\,, &&
4  \left( p  z\right )^2 {\cal T}_6 = 2 T_2 - 2 T_3 - 2 T_4 + 2 T_5 + 2 T_7 + 2 T_8\,,
 \\
4 \, (pz) {\cal T}_7 = T_7 - T_8\,, &&
4  \left( p z \right)^2 {\cal T}_8 = -T_1 + T_2 + T_5 - T_6 + 2 T_7 + 2 T_8 \,.
\\
\end{array}
\renewcommand{\arraystretch}{1.0}  
\eea  
Each distribution amplitude $F = V_i,T_i$ can be 
represented by a Fourier integral 
\beq{fourier}
F(pz) = \int_0^1 dx_3\int_0^{1-x_3}\!\!dx_1 \; e^{-i x_3 (p z)}\, F(x_1,1-x_1-x_3,x_3)\,,
\eeq
where the functions $F(x_i)$ depend on the dimensionless
variables $x_1,x_2=1-x_1-x_3,x_3$, $\,0 < x_i < 1$ which 
correspond to the longitudinal momentum fractions 
carried by the quarks inside the nucleon.  
In difference to the ``calligraphic'' functions ${\cal V}_i(x_i),{\cal T}_i(x_i)$
each of the distribution amplitudes $V_i(x_i),T_i(x_i)$ 
has definite twist, see Table~\ref{tabelle1}, and corresponds 
to the matrix element of a (renormalized) three-quark operator with 
exactly light-like separations $z^2\to 0$, see Table~2 and Appendix C in 
Ref.~\cite{Braun:2000kw} for the details.
    %%%%%%%%%%
    %%%%%%%%%% Begin Table 1
    %%%%%%%%%%
\begin{table}
\renewcommand{\arraystretch}{1.3}
\begin{center}
\begin{tabular}{l|l|l|l}
twist-3  &  twist-4  & twist-5       & twist-6  \\ \hline
$V_1$    & $V_2\;,\;V_3$ & $V_4\;,\;V_5 $& $V_6$ \\ 
$T_1$    & $T_2,T_3,T_7$ & $T_4,T_5,T_8 $& $T_6$ \\ \hline  
\end{tabular}
\end{center}
\caption[]{\sf Twist classification of the distribution amplitudes
in (\ref{opev}).}
\label{tabelle1}
\renewcommand{\arraystretch}{1.0}
\end{table}
     %%%%%%%%%%
     %%%%%%%%%% End Table 1
     %%%%%%%%%%
 The higher-twist distribution 
amplitudes $V_2(x_i),\ldots,V_6(x_i)$ correspond to ``wrong'' components 
of the quark spinors and have different helicity structure compared to 
the leading twist amplitude. For
baryons these ``bad'' components cannot all be traded for gluons as 
in the case of mesons~\cite{Braun:1989iv,Ball:1998sk}. 
They are not all independent, but related to
each other by the exact QCD equations of motion. As a result,  to 
the leading conformal spin accuracy, the five functions 
$V_2(x_i),\ldots,V_6(x_i)$ involve only a single  nonperturbative 
higher twist parameter. 
In the calculations presented
below we use the conformal expansions of higher twist distribution amplitudes
to the next-to-leading order (include ``P-wave'').
This accuracy is consistent with neglecting multiparton components 
with extra gluons (quark-antiquark pairs) that are of yet  higher spin. 
Finally, the invariant functions ${\cal V}_1^M(Pz),{\cal T}^M_1(Pz)$ are twist-5 and can  
be calculated using equations of motion in terms of the nucleon distribution amplitudes~\cite{Braun:2001tj}. 
Explicit expressions for the distribution amplitudes are collected in Appendix~A.   

In the expressions given below we use the following shorthand notations:  
\bea{tildefunction}  
\widetilde{F}(x_3) &=& \int_1^{x_3} \dd x_3' \int_0^{1-x_3'} \dd x_1  
F(x_1,1-x_1-x_3',x_3') \,,  
\nonumber \\  
\widetilde{\widetilde{F}}(x_3) &=& \int_1^{x_3} \dd x_3'\int_1^{x_3'}  
\dd x_3'' \int_0^{1-x_3''} \dd x_1  
F(x_1,1-x_1-x_3'',x_3'')  \,,  
\eea  
where  $F=V_i,T_i$. These functions result from partial integration in $x_3$
which is done in order to eliminate the $1/ p z$ factors that appear in the definition 
of nucleon distribution amplitudes  (\ref{opev}).  After this, 
the $\int d^4 z$ integration becomes trivial. The surface terms sum up to zero.

After a straightforward but tedious calculation we obtain the desired expansions:
\bea{A}
 {\cal A}^{\rm QCD}&=& 4(e_d-e_u)\int\limits_0^1\! dx_3 \Bigg\{
 \frac{x_3}{(x_{3}P-q)^2}\int\limits_0^{1-x_3}\!\!\!dx_1 (V_1-T_1)(x_i)+\frac{x_3m_P^2}{(x_{3}P-q)^4}
 (V_1^{M(d)}-T_1^{M(d)})
\nonumber\\
&&{}+  \frac{x_3^2m_P^2}{(x_{3}P-q)^4}(
 -2 \widetilde V_1 + \widetilde V_2 + \widetilde V_3 + \widetilde V_4 + \widetilde V_5 
 +2 \widetilde T_1 - \widetilde T_2 - \widetilde T_5 - 2\widetilde T_7 - 2\widetilde T_8) 
\nonumber\\
 &&{}+  \frac{2x_3^3m_P^4}{(x_{3}P-q)^6}(
 \wwV_1 -\wwV_2 -\wwV_3-\wwV_4-\wwV_5+\wwV_6 
 -\wwT_1+\wwT_2+\wwT_5-\wwT_6+2\wwT_7+2\wwT_8) 
\nonumber\\
  &&{}+ \left( \frac{x_3m_P^2}{(x_{3}P-q)^4}+ \frac{2x_3m_P^2Q^2}{(x_{3}P-q)^6}\right)(\wwT_2-\wwT_3-\wwT_4+\wwT_5+\wwT_7+\wwT_8)
 \Bigg\},
\eea
\bea{B}
 {\cal B}^{\rm QCD}&=& 4(e_d-e_u)\int\limits_0^1\! dx_3 \Bigg\{
 \frac{x_3 m_P}{(x_{3}P-q)^4}( - \widetilde V_1 + \widetilde V_2 + 
          \widetilde V_3 + \widetilde T_1 - \widetilde T_3 - \widetilde T_7) 
\\
 &+& \frac{2x_3^2m_P^3}{(x_{3}P-q)^6}(
  \wwV_1 -\wwV_2 -\wwV_3-\wwV_4-\wwV_5+\wwV_6  -\wwT_1+\wwT_3+\wwT_4-\wwT_6+ \wwT_7+\wwT_8)
\Bigg\}, 
\nonumber
\eea
\bea{C} 
  {\cal C}^{\rm QCD}&=& 2(e_d-e_u)\int\limits_0^1\! dx_3 \Bigg\{
  \frac{x_3}{(x_{3}P-q)^2}\int\limits_0^{1-x_3}\!\!\!dx_1 (-T_1+V_3)(x_i)
   - \frac{x_3m_P^2}{(x_{3}P-q)^4} T_1^{M(d)}
\nonumber\\
&&{}
  -\frac{1}{(x_{3}P-q)^2}(\widetilde V_1 - \widetilde V_2 - \widetilde V_3)
  +  \frac{Q^2}{(x_{3}P-q)^4}(\widetilde T_1 -\widetilde T_3-\widetilde T_7-\widetilde V_1 +
      \widetilde V_2 + \widetilde V_3)
\nonumber\\
&&{}
+  \frac{x_3^2m_P^2}{(x_{3}P-q)^4}( \widetilde V_4 -  \widetilde V_3 +  \widetilde T_1 -
    \widetilde T_2 +  \widetilde T_3 -  \widetilde T_5 -  \widetilde T_7 - 2  \widetilde T_8)
%- \widetilde S_1+  \widetilde S_2 -  \widetilde P_2 +  \widetilde P_1)
\nonumber\\
&&{}
+\left(\frac{2x_3m_P^2}{(x_{3}P-q)^4}+ \frac{4x_3m_P^2Q^2}{(x_{3}P-q)^6}\right)
(\wwV_1-\wwV_2-\wwV_3-\wwV_4-\wwV_5+\wwV_6)
\nonumber\\
&&{}
+\left(\frac{x_3 m_P^2}{(x_{3}P-q)^4}+ \frac{4x_3m_P^2Q^2}{(x_{3}P-q)^6}\right)
(-\wwT_1+\wwT_2+\wwT_5-\wwT_6+2\wwT_7+2\wwT_8)
\nonumber\\
&&{}- \frac{2x_3m_P^2[Q^2-x_3^2 m_P^2]}{(x_{3}P-q)^6}
(\wwT_2-\wwT_3-\wwT_4+\wwT_5+\wwT_7+\wwT_8)
\Bigg\},
\eea
\bea{D}
   {\cal D}^{\rm QCD}&=& 2\frac{(e_d-e_u)}{m_P}\int\limits_0^1\! dx_3 \Bigg\{
  \frac{1}{(x_{3}P-q)^2}\int\limits_0^{1-x_3}\!\!\!dx_1 V_1(x_i)
   + \frac{m_P^2}{(x_{3}P-q)^4} V_1^{M(d)}
\nonumber\\
&&{}
+  \frac{x_3m_P^2}{(x_{3}P-q)^4}( -3 \widetilde V_1 +  \widetilde V_2 +  2\widetilde V_3
   + \widetilde V_4 +2 \widetilde V_5 +2 \widetilde T_1 -
    \widetilde T_2 -  \widetilde T_5  -  2 \widetilde T_7 - 2  \widetilde T_8)
\nonumber\\
&&{}
 +\frac{4x_3^2m_P^4}{(x_{3}P-q)^6}(
  \wwV_1 -\wwV_2 -\wwV_3-\wwV_4-\wwV_5+\wwV_6 
   -\wwT_1+\wwT_2+\wwT_5-\wwT_6+2\wwT_7+2\wwT_8)
\nonumber\\
&&{}
+\left(\frac{m_P^2}{(x_{3}P-q)^4}+ \frac{2m_P^2[Q^2-x_3^2m_P^2]}{(x_{3}P-q)^6}\right)
(\wwT_2-\wwT_3-\wwT_4+\wwT_5+\wwT_7+\wwT_8)
\Bigg\},
\eea
\bea{E}
{\cal E}^{\rm QCD}&=&
(e_d-e_u)\int\limits_0^1\! dx_3\Bigg\{
\frac{2m_Px_3}{(x_3P-q)^2}\int\limits_0^{1-x_3}\!dx_1 (V_1-V_3)
+\frac{2m_P}{(x_3P-q)^2}(\widetilde{T}_1-\widetilde{T}_3-\widetilde{T}_7)
      \nonumber\\
&&{}+\frac{2x_3m_P^3}{(x_3P-q)^4} V_1^{M(d)}
+\frac{2x_3^2m_P^3}{(x_3P-q)^4}(-\widetilde{V}_1+\widetilde{V}_3+\widetilde{V}_5)
+\frac{2m_P Q^2}{(x_3P-q)^4}(\widetilde{T}_1-\widetilde{T}_3-\widetilde{T}_7)
      \nonumber\\
      &&{}+\frac{2x_3m_P^3}{(x_3P-q)^4}(-\wwT_1+\wwT_3+\wwT_4-\wwT_6+\wwT_7+\wwT_8)
\Bigg\},
\eea
\bea{F}
{\cal F}^{\rm QCD}&=&(e_d-e_u)\int\limits_0^1\! dx_3\Bigg\{
\frac{2}{(x_3P-q)^2}\int\limits_0^{1-x_3}\!dx_1(T_1 +V_1)+\frac{2 m_P^2}{(x_3P-q)^4}(V_1^{M(d)}+T_1^{M(d)})
       \nonumber\\
&&{}-\frac{2x_3m_P^2}{(x_3P-q)^4}(\widetilde{T}_1-\widetilde{T}_3-\widetilde{T}_7
+\widetilde{V}_1-\widetilde{V}_3-\widetilde{V}_5)
       \nonumber\\
&&{}-\frac{2m_P^2}{(x_3P-q)^4}(\wwT_2-\wwT_3-\wwT_4+\wwT_5+\wwT_7+\wwT_8)\Bigg\},
\eea
\bea{G}
      {\cal G}^{\rm QCD}&=& (e_d-e_u)\int\limits_0^1\! dx_3 \Bigg\{
\frac{x_3m_P}{(x_3P-q)^2} \int\limits_0^{1-x_3}\!dx_1(V_3-T_1)(x_i)
+\frac{m_P}{(x_3P-q)^2}(-\widetilde{V}_1+\widetilde{V}_2+\widetilde{V}_3)
\nonumber\\
&&{} -\frac{x_3m_P^3}{(x_3P-q)^4}T_1^{M(d)}
+\frac{m_PQ^2}{(x_3P-q)^4}
(-\widetilde{T}_1+\widetilde{T}_3+\widetilde{T}_7-\widetilde{V}_1+\widetilde{V}_2+\widetilde{V}_3)
\nonumber\\
&&{}+\frac{x_3^2m_P^3}{(x_3P-q)^4}
(\widetilde{T}_1-\widetilde{T}_2+\widetilde{T}_3-\widetilde{T}_5-\widetilde{T}_7-2\widetilde{T}_8
 -\widetilde{V}_3+\widetilde{V}_4)
\nonumber\\
&&{}+\frac{x_3m_P^3}{(x_3P-q)^4}(-\wwT _1+\wwT_2+\wwT_5-\wwT_6+2\wwT_7+2\wwT_8)
\nonumber\\
&&{}+\frac{2x_3m_P^3}{(x_3P-q)^4}(\wwV_1-\wwV_2-\wwV_3-\wwV_4-\wwV_5+\wwV_6)
\nonumber\\
&&{}+\frac{2x_3m_P^3[Q^2+x_3^2m_P^2]}{(x_3P-q)^6}(\wwT_2-\wwT_3-\wwT_4+\wwT_5+\wwT_7+\wwT_8)
\Bigg\},
\eea
\bea{H}
{\cal H}^{\rm QCD}&=&(e_d-e_u)\int\limits_0^1\! dx_3 \Bigg\{
 \frac{-1}{(x_3P-q)^2}\int\limits_0^{1-x_3}\!dx_1(V_1+2T_1)(x_i)-\frac{m_P^2}{(x_3P-q)^4}(V_1^{M(d)}+2T_1^{M(d)})
\nonumber\\
&&{}+\frac{x_3m_P^2}{(x_3P-q)^4}(2\widetilde{T}_1-\widetilde{T}_2-\widetilde{T}_5-2\widetilde{T}_7-2\widetilde{T}_8
+\widetilde{V}_1-\widetilde{V}_2-2\widetilde{V}_3+\widetilde{V}_4)
\nonumber\\
&&{}+\frac{3m_P^2}{(x_3P-q)^4}(\wwT_2-\wwT_3-\wwT_4+\wwT_5+\wwT_7+\wwT_8)
\nonumber\\
&&{}+\frac{2x_3^2m_P^4}{(x_3P-q)^6}(\wwT_2-\wwT_3-\wwT_4+\wwT_5+\wwT_7+\wwT_8)
\nonumber\\
&&{}+\frac{2m_P^2 Q^2}{(x_3P-q)^6}(\wwT_2-\wwT_3-\wwT_4+\wwT_5+\wwT_7+\wwT_8)\Bigg\},
\eea
\bea{I}
      {\cal I}^{\rm QCD}&=&(e_d-e_u)Q^2\int\limits_0^1\! dx_3 \Bigg\{
\frac{1}{(x_3P-q)^2}\int\limits_0^{1-x_3}\!dx_1(V_1-2T_1)(x_i)+\frac{m_P^2}{(x_3P-q)^4}(V_1^{M(d)}\!-2T_1^{M(d)})
\nonumber\\
&&{}-\frac{x_3m_P^2}{(x_3P-q)^4}(
{-6\widetilde{T}_1+3\widetilde{T}_2+3\widetilde{T}_5+6\widetilde{T}_7+6\widetilde{T}_8}
    +5\widetilde{V}_1-\widetilde{V}_2-2\widetilde{V}_3-3\widetilde{V}_4-4\widetilde{V}_5)
\nonumber\\
&&{}+\frac{5m_P^2}{(x_3P-q)^4}(\wwT_2-\wwT_3-\wwT_4+\wwT_5+\wwT_7+\wwT_8)
\nonumber\\
&&{}-\frac{2x_3^2m_P^4}{(x_3P-q)^6}(4\wwT_1-3\wwT_2-\wwT_3-\wwT_4-3\wwT_5+4\wwT_6-7\wwT_7-7\wwT_8)
\nonumber\\
&&{}-\frac{8x_3^2m_P^4}{(x_3P-q)^6}(-\wwV_1+\wwV_2+\wwV_3+\wwV_4+\wwV_5-\wwV_6)
\nonumber\\
&&{}-\frac{6m_P^2Q^2}{(x_3P-q)^6}(-\wwT_2+\wwT_3+\wwT_4-\wwT_5-\wwT_7-\wwT_8)\Bigg\},
\eea
\bea{J}
{\cal J}^{\rm QCD}&=&(e_d-e_u)\int\limits_0^1\! dx_3 \Bigg\{
\frac{x_3m_P}{(x_3P-q)^2} \int\limits_0^{1-x_3}\! dx_1 (V_3-T_1)(x_i)
+\frac{m_P}{(x_3P-q)^2}(-\widetilde{V}_1+\widetilde{V}_2+\widetilde{V}_3)
\nonumber\\
&&{}-\frac{x_3m_P^3}{(x_3P-q)^4}T_1^{M(d)}
+\frac{m_P Q^2}{(x_3P-q)^4}
(3\widetilde{T}_1-3\widetilde{T}_3-3\widetilde{T}_7-\widetilde{V}_1+\widetilde{V}_2+\widetilde{V}_3)
\nonumber\\
&&{}+\frac{x_3m_P^3}{(x_3P-q)^4}(-\wwT_1+\wwT_2+\wwT_5-\wwT_6+2\wwT_7+2\wwT_8)
\nonumber\\
&&{}+\frac{2x_3m_P^3}{(x_3P-q)^4}(\wwV_1-\wwV_2-\wwV_3-\wwV_4-\wwV_5+\wwV_6)
\nonumber\\
&&{}+\frac{x_3^2m_P^3}{(x_3P-q)^4}(\widetilde{T}_1-\widetilde{T}_2+\widetilde{T}_3-\widetilde{T}_5-\widetilde{T}_7-2\widetilde{T}_8-\widetilde{V}_3+\widetilde{V}_4)
\nonumber\\
&&{}+\frac{2x_3^3m_P^5}{(x_3P-q)^6}(\wwT_2-\wwT_3-\wwT_4+\wwT_5+\wwT_7+\wwT_8)
\nonumber\\
&&{}+\frac{x_3m_P^3Q^2}{(x_3P-q)^6}(-8\wwT_1+2\wwT_2+6\wwT_3+6\wwT_4+2\wwT_5-8\wwT_6+10\wwT_7+10\wwT_8)
\nonumber\\
&&{}+\frac{8x_3m_P^3Q^2}{(x_3P-q)^6}(\wwV_1-\wwV_2-\wwV_3-\wwV_4-\wwV_5+\wwV_6)\Bigg\}.
\eea

The Borel transformation and the continuum subtraction  
are performed by using the following substitution rules:  
\bea{borel0}  
\int \dd x \frac{\varrho(x) }{(q-x  P)^2} &=&  
- \int_0^1 \frac{\dd x }{x} \frac{\varrho(x)}{(s - {P'}^2)}  
\nonumber \\  &\to&  
-   \int_{x_0}^1 \frac{\dd x}{x} \varrho(x)  
\exp{%\displaystyle  
\left( - \frac{\bar x Q^2}{x M_B^2}  - \frac{\bar x m_\Delta^2}{M_B^2}\right)}\, ,    
\nonumber\\
\int \dd x \frac{\varrho(x) }{(q-x  P)^4} &=&  
\int_0^1 \frac{\dd x }{x^2} \frac{\varrho(x)}{(s - {P'}^2)^2}  
\nonumber \\  &\to&  
\frac{1}{M_B^2} \int_{x_0}^1 \frac{\dd x}{x^2} \varrho(x)  
\exp{%\displaystyle  
\left( - \frac{\bar x Q^2}{x M_B^2}  - \frac{\bar x m_\Delta^2}{M_B^2}\right)}    
+  
\frac{\varrho(x_0)\,e^{-s_0 /M_B^2} }{Q^2 + x_0^2 m_\Delta^2} \, ,    
\nonumber\\
\int \dd x \frac{\varrho(x) }{(q-x  P)^6} &=&  
- \int_0^1 \frac{\dd x }{x^3} \frac{\varrho(x)}{(s - {P'}^2 )^3}  
\nonumber \\  &\to & -  
\frac{1}{2 M_B^4} \int_{x_0}^1 \frac{\dd x}{x^3} \varrho(x)  
\exp{  
\left( - \frac{\bar x Q^2}{x M_B^2} - \frac{\bar x m_\Delta^2}{M_B^2}  
\right)}  
- \frac12  
\frac{\varrho(x_0)\,e^{-s_0 /M_B^2} }{x_0\left(Q^2 + x_0^2 m_\Delta^2\right) M_B^2}  
\nonumber \\  &&  
+ \frac12  \frac{x_0^2}{Q^2 + x_0^2 m_\Delta^2} \left[\frac{d}{dx_0}  
\frac{\varrho(x_0)}{x_0\left(Q^2 + x_0^2 m_\Delta^2\right)} \right]  
\,e^{-s_0 /M_B^2} \, \nonumber \\  
\eea  
where $M_B$ is the Borel parameter, 
$s = \frac{1-x}{x} Q^2  + (1-x) m_\Delta^2$ and 
$x_0$ is the solution of the corresponding quadratic equation for $s = s_0$:
\bea{x0}
   x_0 &=&\bigg[ \sqrt{(Q^2+s_0-m_\Delta^2)^2+ 4 m_\Delta^2 Q^2}-(Q^2+s_0-m_\Delta^2)\bigg]
   /(2m_\Delta^2)\,.
\eea  
The contributions $\sim e^{-s_0 /M_B^2}$ in \Gl{borel0}
correspond to the ``surface terms'' arising from  successive  
partial integrations to reduce the power in the denominators  
$(q - x P)^{2N} = (s  - {P'}^2 )^{2N} (-x)^{2N}$  
with $N > 1$ to the usual dispersion representation with the denominator  
$\sim (s - {P'}^2 )$. Without continuum subtraction, i.e. in the 
limit  $s_0 \to \infty$ these terms vanish.  

In addition, in the hadronic representation for the same correlation 
functions (\ref{lhs}) one has to make the substitution
\beq{borel5}
 \frac{1}{m_\Delta^2-P'^2} \to e^{-m_\Delta^2/M_B^2}.
\eeq

%%%%%%%%%%%%%%%%%
\setcounter{equation}{0} \section{Light-cone Sum Rules}
%%%%%%%%%%%%%%%%%

Our general strategy is as follows.
At the first step, we analyze the two light-cone sum rules
that are obtained from the combinations of the correlation functions
that are free from contributions of 
negative parity spin-1/2 resonances, cf. (\ref{free12}):
\bea{freeSR}
   {\mathbb B}\Big[{\cal G}^{\rm QCD} - {\cal J}^{\rm QCD} + Q^2 {\cal B}^{\rm QCD}
   \Big](M_B^2) &=& 
- \frac{\lambda_\Delta e^{-m_\Delta^2/M_B^2}}{(2\pi)^2} Q^2 G_2(Q^2)\,,
\\     
   {\mathbb B}\Big[{\cal I}^{\rm QCD} + Q^2\,{\cal H}^{\rm QCD} -  Q^2\,{\cal A}^{\rm QCD}
    \Big](M_B^2) &=&
- \frac{\lambda_\Delta e^{-m_\Delta^2/M_B^2}}{(2\pi)^2}Q^2\big[2 G_1(Q^2)+(m_\Delta-m_P)G_2(Q^2)\big].
\nonumber
\eea
Here ${\mathbb B}\big[\ldots\big](M_B^2)$ stands for the Borel transform with respect to $P'^2$.
{}From these sum rules we extract $G_1$ and $G_2$ form factors and also $G_M-G_E\sim G_1$.
At the second step, we determine the form factor $G_3$ from the leading twist 
sum rule
\beq{B-SR}
    {\mathbb B}\Big[{\cal B}^{\rm QCD}\Big](M_B^2) =
 - \frac{\lambda_\Delta e^{-m_\Delta^2/M_B^2}}{(2\pi)^2M_B^2}
\frac{1}{3m_{\Delta}^{2}}
\Big[-2m_{P}G_{1}+[m_{\Delta}(m_{\Delta}-m_{P})-Q^{2}]G_{2}+2Q^{2}G_{3}\Big].
\eeq  
The rationale for choosing this structure is that in the sum rule for ${\cal A}$ the contribution 
of interest is multiplied by a small factor $m_\Delta-m_P$, cf.~(\ref{lhs}).
{}Finally, we rewrite our results in terms  of $G_M$, $G_E$ and $G_C$.

\subsection{Asymptotic expansion}

As an example, consider the contribution of the leading-twist nucleon distribution amplitudes to the LCSR 
for $G_2(Q^2)$, the first equation in (\ref{freeSR}). Putting everything together, one obtains to this 
accuracy
\bea{example1}
 {\cal G}^{\rm QCD} - {\cal J}^{\rm QCD} + Q^2 {\cal B}^{\rm QCD} &=&
 (e_d-e_u)Q^2\int\limits_0^1\! dx_3 \Bigg\{
  -\frac{4m_P }{(x_3P-q)^4}\widetilde{T}_1  
+\frac{4x_3m_P }{(x_3P-q)^4}(\widetilde{T}_1-\widetilde{V}_1)
\nonumber\\&&{}
+\frac{x_3\bar x_3 m_P^3}{(x_3P-q)^6}(\wwT_1-\wwV_1)    
                                        \Bigg\} 
+{\cal O}(\mbox{\rm twist-4})\,.
\eea
The main effect of the continuum subtraction and/or Borel transformation 
is that the integration over the momentum fraction $x_3$ 
gets restricted to a narrow interval of $1 > x_3 > x_0$ with $1-x_0 =s_0/Q^2 \ll 1$. As the result, power 
counting of  $1-x_3$ factors in the integrand translates into the power counting in $1/Q^2$.
The behavior of the nucleon distribution amplitudes close to the end point $x_3\to 1$ is governed by 
conformal symmetry \cite{Braun:2003rp}. To the next-to-leading order accuracy in the expansion over the 
conformal spin one obtains~\cite{Chernyak:1984bm,Braun:2000kw}
\bea{conf1}
 V_1(x_1,x_2,x_3) &=& 120 f_N x_1 x_2 x_3 \Big[1+\frac72(1-3V_1^d)(1-3x_3)+\ldots\Big],
\nonumber\\
 T_1(x_1,x_2,x_3) &=& 120 f_N x_1 x_2 x_3 \Big[1+ \frac74(3 A_1^u+3V_1^d-1)(1-3x_3)+\ldots\Big]. 
\eea 
Here $A_1^u$ and $V_1^d$ are scale-dependent parameters that characterize the deviation of the 
distribution amplitude from its asymptotic form at $Q^2\to\infty$. For the asymptotic distribution
amplitude $V_1^d=1/3$ and $A_1^u =0$; the Chernyak-Zhitnitsky (CZ) model \cite{Chernyak:1984bm} corresponds to 
$V_1^d=0.23$ and $A_1^u =0.38$ at a low scale of a few hundred MeV.  
A simple calculation shows that in the end-point region $\widetilde{V}_1\sim \widetilde{T}_1 \sim (1-x_3)^4$  
and $\wwV_1\sim \wwT_1\sim (1-x_3)^5$.

After some algebra we obtain the contribution of leading-twist nucleon distribution amplitudes to    
the light-cone sum rule (\ref{freeSR}) for $G_2(Q^2)$ in the $Q^2\to \infty$ limit:
\bea{example2}
 \frac{\lambda_\Delta e^{-m_\Delta^2/M_B^2}}{(2\pi)^2} G^{\rm tw-3}_2(Q^2) &=&
  \frac{4(e_u-e_d)m_P}{Q^4}\int\limits_0^{s_0}ds\, e^{-s/M_B^2}
\int\limits_0^{s/Q^2} dx_1\, V_1(x_1,s/Q^2-x_1,1-s/Q^2)
\nonumber\\
&=&\frac{80 f_N m_P}{Q^{10}}\big[1+7(3V_1^d-1)\big]\int\limits_0^{s_0} ds\, s^3 \, e^{-s/M_B^2}
+{\cal O}\left(\frac{1}{Q^{12}}\right).
\eea
The term in $\sim (3V_1^d-1)$ corresponds to the contribution of subleading conformal spin, 
cf.~(\ref{conf1}). For the CZ-model this correction is very large, 
$\big[1+7(3V_1^d-1)\big]_{\rm CZ} \simeq -1.17$.
It turns out, however, that the contribution of the leading-twist distribution amplitudes 
is only subleading for large $Q^2$, and the true large-$Q^2$ asymptotics of the sum rules 
is determined by the contribution of higher-twist amplitudes, in particular those corresponding to different 
helicity structures (compared to leading twist). Using full expressions, i.e. 
taking into account the terms $\sim V_2\ldots V_6$ and $\sim T_2\ldots T_6$, 
we obtain
\bea{asymG2}
  \frac{\lambda_\Delta e^{-m_\Delta^2/M_B^2}}{(2\pi)^2} G_2(Q^2) &=&
   \frac{8 m_P}{Q^8}\big[5 f_N+ 3 \lambda_1]\int\limits_0^{s_0} ds\, s^2 \, e^{-s/M_B^2}
  +{\cal O}\left(\frac{1}{Q^{10}}\right),
\eea
\bea{asymG12}
 \frac{\lambda_\Delta e^{-m_\Delta^2/M_B^2}}{(2\pi)^2}\big[2 G_1(Q^2)+(m_\Delta-m_P)G_2(Q^2)\big]
&=&
    \frac{8}{Q^8}\Bigg\{
  10 f_N \int\limits_0^{s_0} ds\, s^3 \, e^{-s/M_B^2}
\nonumber\\&&
 \hspace*{-4.2cm}{}- m_P^2 \Big[\frac{89}{3}f_N+ 7 \lambda_1\Big] \int\limits_0^{s_0} ds\, s^2 \, e^{-s/M_B^2}\Bigg\}
  +{\cal O}\left(\frac{1}{Q^{10}}\right),
\eea
\bea{asymG123}
 \frac{\lambda_\Delta e^{-m_\Delta^2/M_B^2}}{(2\pi)^2}\big[2G_3(Q^2) - G_2(Q^2)\big]
&=&
    \frac{8m_\Delta^2 m_P}{Q^{10}}\Bigg\{
  \big[75 f_N +9 \lambda_1\big]\int\limits_0^{s_0} ds\, s^2 \, e^{-s/M_B^2}
\nonumber\\
&& \hspace*{-2.2cm}{}+
 m_P^2 \Big[34 f_N+ \frac{24}{5} \lambda_1\Big] \int\limits_0^{s_0} ds\, s \, e^{-s/M_B^2}\Bigg\}
  +{\cal O}\left(\frac{1}{Q^{12}}\right),
\eea
from the three sum rules in Eqs.~(\ref{freeSR}), (\ref{B-SR}), respectively. 
{}For simplicity, only the contributions of asymptotic distribution amplitudes are shown, see 
Appendix~B. 

To avoid misunderstanding, note that these expressions correspond to the ``soft'' contribution 
to the form factors that does not involve any hard gluon exchanges and correspond, loosely speaking, 
to the first term in the expansion in Eq.~(\ref{schema}). In order to observe the true asymptotic behavior 
one has to calculate radiative corrections to the sum rules, cf. \cite{Braun:1999uj}.
In this way, the hard gluon exchange contribution considered in \cite{Carlson:1988gt}  appears to be 
a part of the two-loop ${\cal O}(\alpha^2_s)$ correction.
The corresponding calculation goes beyond the scope  of the present work. 
In spite of being enhanced asymptotically 
by  powers of the momentum transfer, the radiative corrections are  accompanied 
by increasing powers of the (small) QCD coupling at a scale at least that of the Borel parameter, and 
we expect that for moderate momentum transfers in the range $1< Q^2 < 10$~GeV$^2$ their contribution is
numerically less important.

%
%%%%%%%%%%%%%%%%%%     FIGURE 1          %%%%%%%%%%%%%%%%%%%%%%%%%%%%
\begin{figure}[ht]
\centerline{\epsfysize6.5cm\epsffile{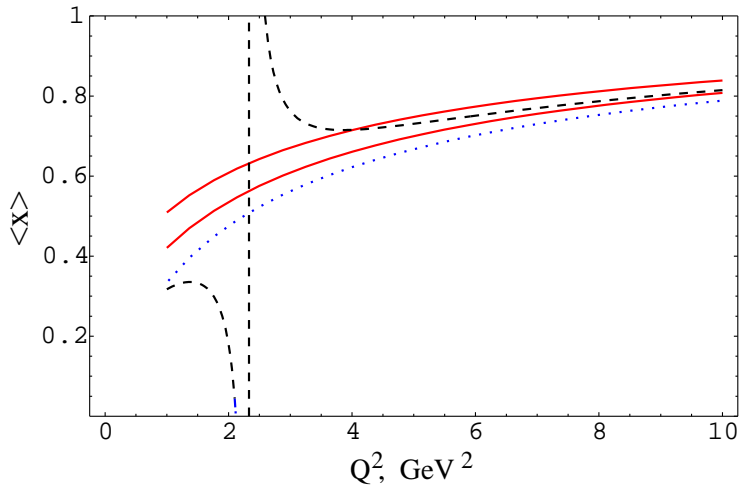}}
\caption[]{\small
The average value $\langle x\rangle$ of the momentum fraction $x_3$ in the sum rules 
(\ref{freeSR}), (\ref{B-SR}) as a function of the momentum transfer $Q^2$. 
The lower and the upper solid curves correspond to the first and the second sum rule 
in (\ref{freeSR}), respectively. The dashed curve corresponds to (\ref{B-SR}). 
The value of the threshold $x_0$ (\ref{x0}) is plotted by dots for comparison.} 
\label{fig:x0}
\end{figure}
%%%%%%%%%%%%%%%%%%%%%%%%%%%%%%%%%%%%%%%%%%%%%%%%%%%%%%%%%%%%%%%%%%%%%%
%

We conclude that the leading-order LCSR calculation predicts a universal $1/Q^8$ falloff of soft contributions 
to all the three form factors $G_1$, $G_2$, $G_3$, and to this accuracy $G_2 = 2G_3$.
This translates to the asymptotic behavior of the soft terms 
$G_M \sim 1/Q^6$, $G_M \sim 1/Q^6$ and $G_Q \sim 1/Q^8$, for the magnetic, electric and quadrupole form factors 
defined in Eq.~(\ref{MEQ}), respectively.    
In agreement with the common wisdom, soft contributions in the light-cone sum rules
arise from the integration regions where the quark interacting with 
the virtual photon carries almost all hadron momentum, alias the corresponding momentum fraction $x\to 1$. 
To illustrate this feature, we have plotted in Fig.~\ref{fig:x0} the average value $\langle x\rangle$ 
of the momentum fraction $x_3$ in the integrals in the sum rules (\ref{freeSR}), (\ref{B-SR}).  
The main effect here is that 
the integration region in the momentum fraction gets restricted 
to a narrow interval $x_0 < x < 1$ where $x_0$ is given by Eq.~(\ref{x0}). For asymptotically large $Q^2$ 
one finds $x_0\simeq 1-s_0/Q^2$ and the integration region shrinks to the end-point. For realistic values 
of $Q^2$ in the range $1-10$~GeV$^2$ the average 
 $\langle x\rangle$  grows slowly  and one finds  very similar values   
for all the three sum rules in question. [The irregular behavior that is seen for the sum rule 
(\ref{B-SR}) around $Q^2\sim 2$~GeV$^2$ is due to accidental vanishing of the denominator involved 
in taking the average.]

\subsection{Numerical analysis}

The numerical results presented below are obtained using two models of the nucleon distribution
amplitudes \cite{Braun:2000kw}. The first model corresponds to asymptotic distribution amplitudes
of all twists. The corresponding expressions are collected in Appendix~B. The second model 
corresponds to taking into account the corrections to the asymptotic distribution amplitudes
that are due to the next-to-leading conformal spin (``P-waves'') with parameters estimated using 
QCD sum rules. For the leading twist distribution amplitude this choice corresponds 
to a simplified version of the Chernyak-Zhitnitsky (CZ) model: 
we have truncated the original CZ expressions \cite{Chernyak:1984bm} leaving out contributions 
of the next-to-next-to-leading conformal spin operators (``D-waves'') in order to simplify the calculation 
of nucleon mass corrections, see \cite{Braun:2000kw} for details. The explicit expressions of the
distribution amplitudes including P-wave corrections are rather cumbersome, 
they can be found in \cite{Braun:2000kw,Lenz05}. 

In addition, for the numerical evaluation of the sum rules we have to specify 
the values of the $\Delta$-coupling $\lambda_\Delta$, the continuum threshold $s_0$ and 
the Borel parameter $M_B^2$. The usual strategy is to determine all of them from the simplest 
QCD sum rule involving the $\Delta$ resonance \cite{Belyaev:1982sa}, see Appendix C. 
We obtain from this sum rule $s_0\simeq3$~GeV$^2$ and   
$\lambda_\Delta \simeq 2.0$~GeV$^3$. Both numbers are somewhat lower that the ones obtained 
in \cite{Belyaev:1982sa} because we use the experimental input value of 
the mass of the $\Delta$-resonance $m_\Delta=1.23$~GeV instead of trying to find it from the sum rule.      

%
%%%%%%%%%%%%%%%%%%     FIGURE 2          %%%%%%%%%%%%%%%%%%%%%%%%%%%%
\begin{figure}[t]
\centerline{\epsfysize6.5cm\epsffile{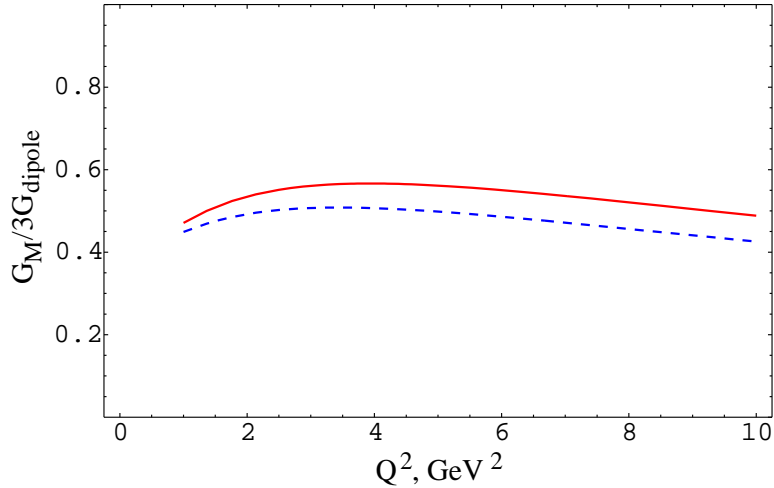}}
\caption[]{\small
The Borel parameter dependence of the ratio $G_M/3 G_{\rm dipole}$ where $G_{\rm dipole} = 1/(1+Q^2/0.71)^2$.
The magnetic form factor $G_M$ is defined in (\ref{MEQ}). The solid and the dashed curves correspond 
to the calculation using $M_B^2 = 2.5$~GeV$^2$ and $M_B^2 = 1.5$~GeV$^2$, respectively.
In both cases the asymptotic nucleon distribution amplitudes of the 
leading and higher twists are used.}
\label{fig:stability}
\end{figure}
%%%%%%%%%%%%%%%%%%%%%%%%%%%%%%%%%%%%%%%%%%%%%%%%%%%%%%%%%%%%%%%%%%%%%%
%
A suitable range of Borel parameters for the LCSR can be obtained as follows. On the one hand, 
$M_B^2$ has to be small enough in order to guarantee sufficient suppression of higher mass resonances
and the continuum in the hadronic representation for the correlation function. This is the same criterium 
that is applied to the Belyaev-Ioffe sum rule (\ref{BI-SR}) for the coupling $\lambda_\Delta$, hence we 
would want to take $M_B^2$ as close as possible to 1.5~GeV$^2$ which is the minimum value at 
which the stability in (\ref{BI-SR}) sets in. 
On the other hand, the Borel parameter in the LCSRs has to be large enough to guarantee convergence of the twist
expansion in the QCD calculation.
Note that for a fixed value of the momentum fraction $x$ the light-cone expansion of the relevant correlation 
functions goes generically in powers of $1/(q-xP)^2$ which translates to the expansion in powers of 
$1/(x M^2_B)$ after the Borel transformation. The true expansion parameter is therefore 
of order $\sim 1/(\langle x\rangle M^2_B)$ where $\langle x\rangle$ is   
the average value of the momentum fraction in the corresponding integrals, rather than $1/M_B^2$ itself.
{}In the range $1 < Q^2 < 10$~GeV$^2$ we find $0.4 < \langle x\rangle <  0.8$ (see Fig.~\ref{fig:x0}) so
that preferred values of $M_B^2$ in LCSRs appear from this side to be a factor 1.2--2.5 larger compared 
to the two-point sum rule Borel parameter $M^2$. 
Note that  for fixed $M^2$ there is effectively a bias towards  using larger values of $M_B^2$ 
with decreasing  $Q^2$. All in all, it appears that the interval 
$1.5 < M^2_B < 2.5$~GeV$^2$ presents a reasonable choice. The stability of the LCSRs 
in this range turns out to be better than 10--15\%, see Fig.~\ref{fig:stability}
for an example. In what follows we use the fixed value of the Borel parameter $M^2_B=2$~GeV$^2$ 
in the center of this fiducial interval.

%
%%%%%%%%%%%%%%%%%%     FIGURE 3          %%%%%%%%%%%%%%%%%%%%%%%%%%%%
\begin{figure}[t]
\centerline{\epsfysize7cm\epsffile{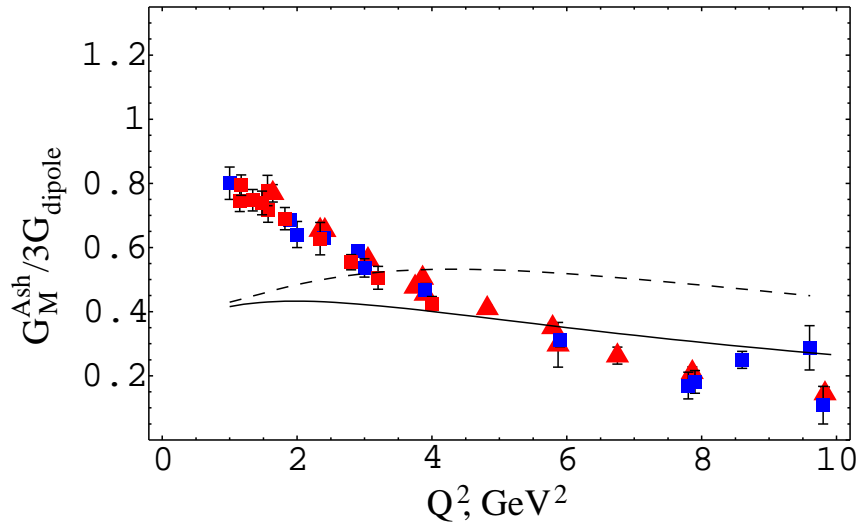}}
\caption[]{\small
The ratio $G_M^{\rm Ash}/3 G_{\rm dipole}$ where $G_{\rm dipole} = 1/(1+Q^2/0.71)^2$.
The form factor $G_M^{\rm Ash}$ is defined in (\ref{GT}). The solid and the dashed curves correspond 
to the calculation using the asymptotic and the 
QCD sum rules motivated  nucleon distribution amplitudes of the leading and higher twists, respectively.
The data points are from \cite{Stoler:1993yk} (blue squares) \cite{Stuart:1996zs} (red triangles)
and \cite{Bartel:1968tw,Alder:1972di,Stein:1975yy,Foster:1983kn,Frolov:1998pw} (red squares).}
\label{fig:GMAsh}
\end{figure}
%%%%%%%%%%%%%%%%%%%%%%%%%%%%%%%%%%%%%%%%%%%%%%%%%%%%%%%%%%%%%%%%%%%%%%
%
%
%%%%%%%%%%%%%%%%%%     FIGURE 4          %%%%%%%%%%%%%%%%%%%%%%%%%%%%
\begin{figure}[t]
\centerline{\epsfysize7cm\epsffile{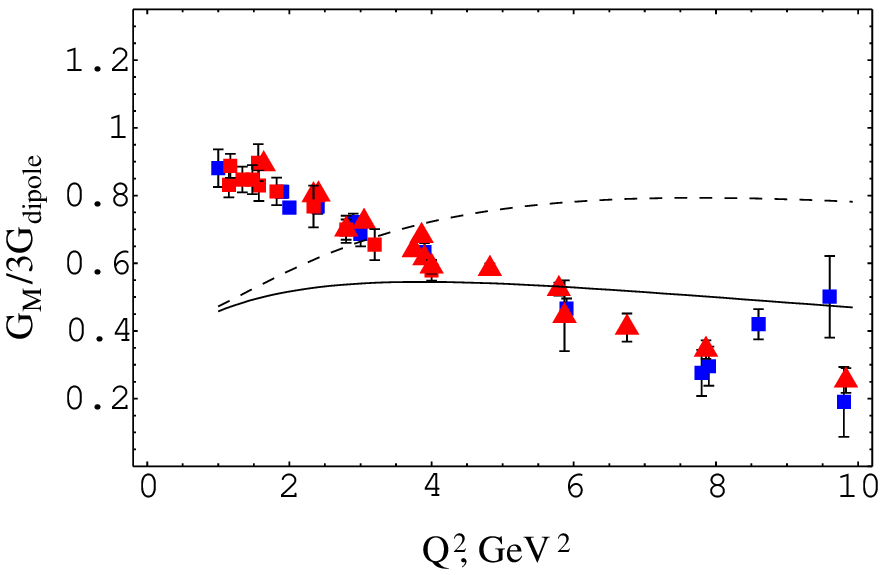}}
\caption[]{\small
The ratio $G_M/3 G_{\rm dipole}$ where $G_{\rm dipole} = 1/(1+Q^2/0.71)^2$.
The magnetic form factor $G_M$ is defined in (\ref{MEQ}). 
The identification of the curves and data points is the same as in Fig.~\ref{fig:GMAsh}.}
\label{fig:GM}
\end{figure}
%%%%%%%%%%%%%%%%%%%%%%%%%%%%%%%%%%%%%%%%%%%%%%%%%%%%%%%%%%%%%%%%%%%%%%
%

%
%%%%%%%%%%%%%%%%%%     FIGURE 5          %%%%%%%%%%%%%%%%%%%%%%%%%%%%
\begin{figure}[ht]
\centerline{\epsfysize7cm\epsffile{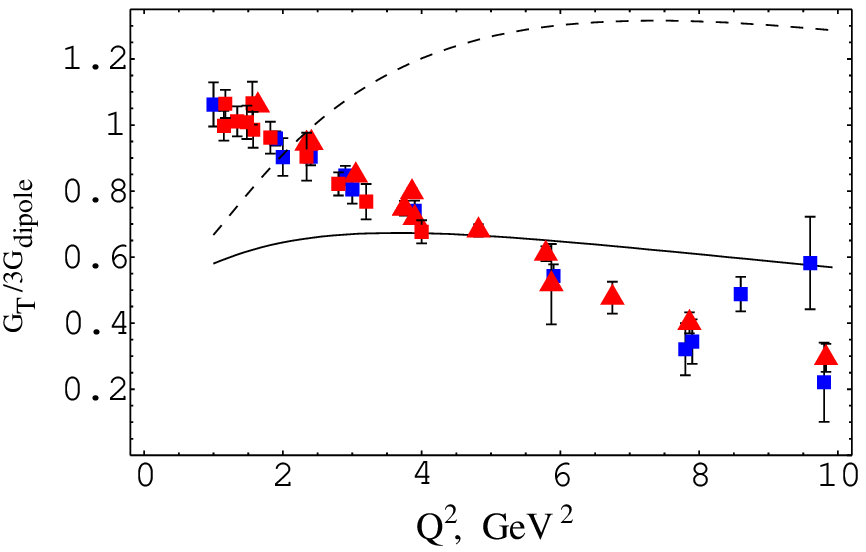}}
\caption[]{\small
The ratio $G_T/3 G_{\rm dipole}$ where $G_{\rm dipole} = 1/(1+Q^2/0.71)^2$.
The form factor $G_T$ is defined in (\ref{GT}). 
The identification of the curves and data points is the same as in Fig.~\ref{fig:GMAsh}.}
\label{fig:GT}
\end{figure}
%%%%%%%%%%%%%%%%%%%%%%%%%%%%%%%%%%%%%%%%%%%%%%%%%%%%%%%%%%%%%%%%%%%%%%
%

The results for the magnetic transition form factor are shown in Fig.~\ref{fig:GMAsh}, Fig.~\ref{fig:GM} 
and Fig.~\ref{fig:GT} using three different ways to present the data (in terms of $G_M^{\rm Ash}$, $G_M$ and 
$G_T$) accepted in the experimental papers. The difference between $G_M^{\rm Ash}$ and 
 $G_M$ is in a kinematical factor only, while $G_T$ includes in addition the contribution of the electric
form factor $G_E$ which is negligible. 
We observe that the calculation with asymptotic distribution amplitudes (solid curves) 
is much closer to the data so that large deviations from the asymptotic shape of distribution amplitudes 
suggested by QCD sum rule calculations are 
disfavored.  A similar conclusion was reached 
in Refs.~\cite{Braun:2001tj,Lenz:2003tq} from the LCSR analysis of 
the electromagnetic nucleon form factors. 

\begin{figure}[ht]
\centerline{\epsfysize5cm\epsffile{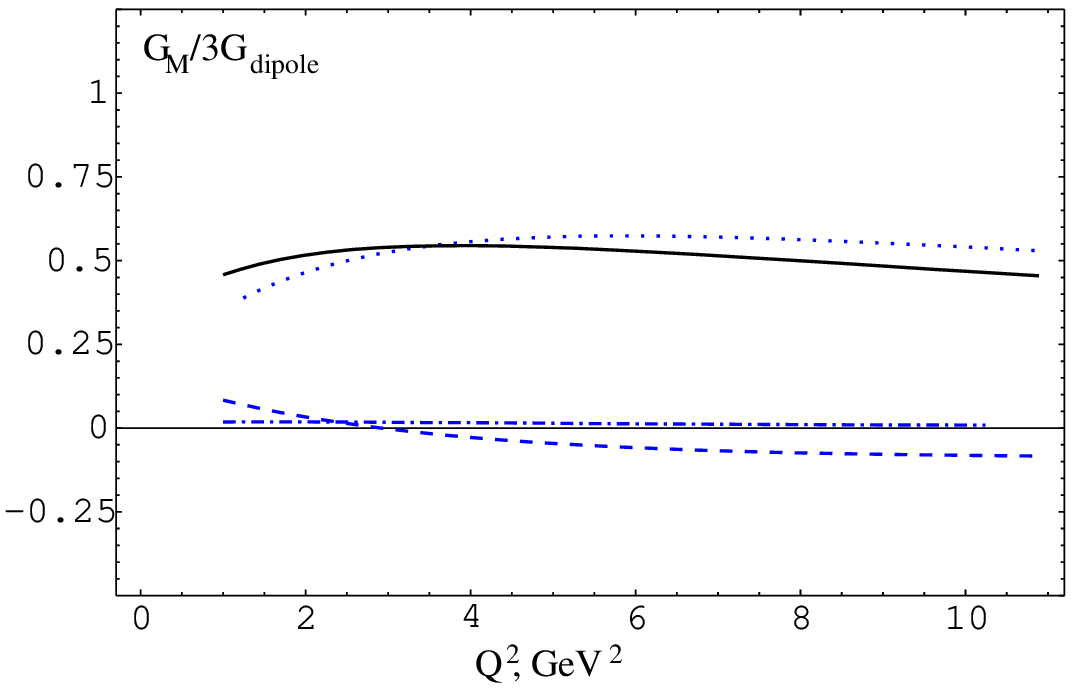},\epsfysize5cm\epsffile{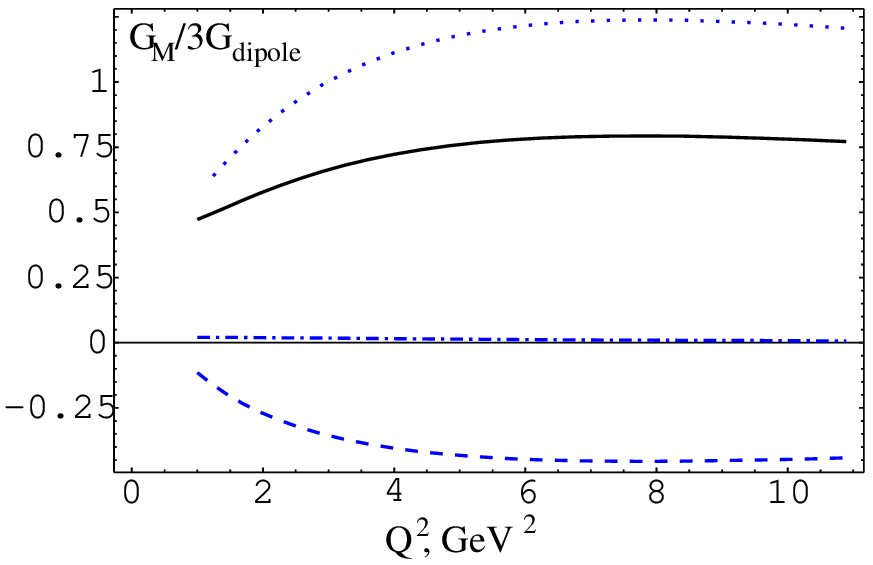}}
\caption[]{\small Separate contributions to the light-cone sum rule results (solid curves) for
the ratio $G_M/3 G_{\rm dipole}$ where $G_{\rm dipole} = 1/(1+Q^2/0.71)^2$.
The terms in $f_N$, $\lambda_1$ and $\lambda_2$ are shown by the dashed, dotted and dash-dotted 
curves, respectively. The results on the left panel correspond to the  asymptotic distribution 
amplitudes and the ones on the right panel are obtained including a CZ-like model, see text.}
\label{fig:decompose}
\end{figure}

It turns out that the sum rules for $G_M$  are dominated by contributions of subleading twist-4. 
To illustrate this issue, we have plotted in Fig.~\ref{fig:decompose} separate contributions to 
the sum rule for the ratio $G_M/(3G_{dipole})$ that come from the  nonperturbative matrix 
element of twist three $\sim f_N$ (dashed curves) and the two existing
matrix elements of twist four: terms in $\lambda_1$ (dotted curves) and $\lambda_2$ (dash-dotted curves), 
cf. Appendix B. The full result is given by the sum of all three contributions and is shown by the solid curves.
We see that contributions $\sim \lambda_2$ are numerically very small, the contributions $\sim \lambda_1$ 
are the dominant ones and remain roughly constant in the considered $Q^2$ range, whereas the contributions 
$\sim f_N$ have a stronger $Q^2$ dependence and enter with an opposite sign. QCD sum rule motivated 
corrections to the asymptotic distribution amplitudes generally tend to increase both leading- and higher-twist 
contributions, so that the cancellation becomes more pronounced.
Note that the uncertainties in 
nonperturbative parameters $f_N,\lambda_1,\lambda_2$ presently are at the level of 20-30\%. Larger leading twist contributions
$\sim f_N$ would yield a steeper falloff of the form factor which is favored by the data. The same observation
was made in \cite{Braun:2001tj} for the case of the electromagnetic form factors of the nucleon.  
Note that the dominant contributions $\sim \lambda_1$ correspond to the operators that include a ``minus'' component 
of one of the valence quark fields. They do not have a simple partonic interpretation in terms of quark parton 
amplitudes at small transverse separation but rather involve the orbital angular momentum, see 
\cite{Ji:2002xn,Ji:2003yj} for the relevant formalism and  discussion.

%
%%%%%%%%%%%%%%%%%%     FIGURE 5a          %%%%%%%%%%%%%%%%%%%%%%%%%%%%
%%%%%%%%%%%%%%%%%%%%%%%%%%%%%%%%%%%%%%%%%%%%%%%%%%%%%%%%%%%%%%%%%%%%%%
%

\begin{figure}[ht]
\centerline{\epsfysize7cm\epsffile{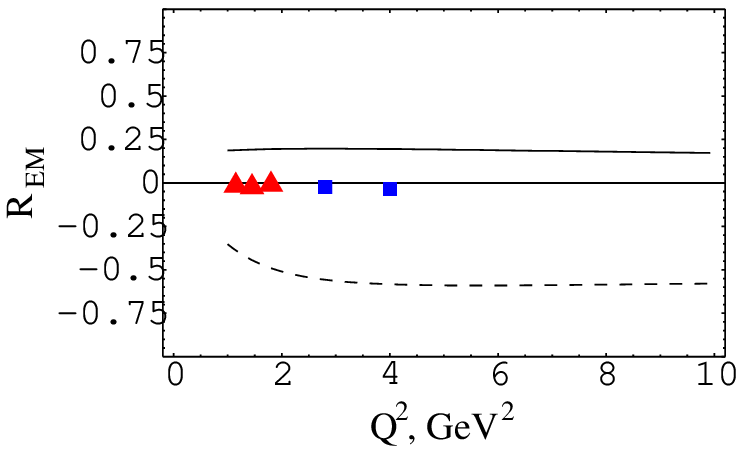}}
\caption[]{\small
The ratio $R_{\rm EM}(Q^2)= - G_E/G_M$ of the electric and the magnetic form factors, 
cf.~(\ref{GT}).
The solid and the dashed curves correspond 
to the calculation using the asymptotic and the 
QCD sum rules motivated  nucleon distribution amplitudes of the leading and higher twists, respectively.
The data points are from \cite{Joo:2001tw} (red squares) and \cite{Frolov:1998pw} (blue squares).}
\label{fig:GE}
\end{figure}

\begin{figure}[ht]
\centerline{\epsfysize7cm\epsffile{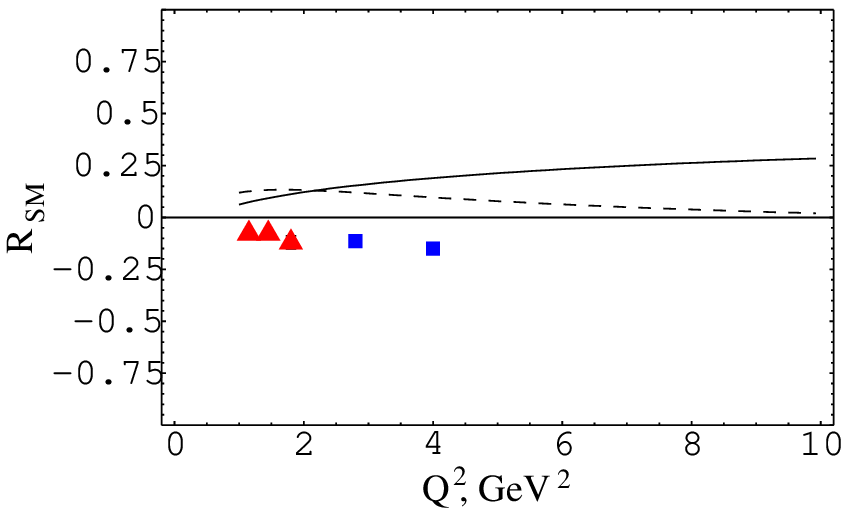}}
\caption[]{\small
The ratio $R_{SM}\sim -G_C/G_M$ of the quadrupole and the magnetic form factors, 
cf.~(\ref{GT}).
The solid and the dashed curves correspond 
to the calculation using the asymptotic and the 
QCD sum rules motivated  nucleon distribution amplitudes of the leading and higher twists, respectively.
The data points are from \cite{Joo:2001tw} (red squares) and \cite{Frolov:1998pw} (blue squares).}
\label{fig:GC}
\end{figure}

The results for the electric form factor $G_E$ and quadrupole form factor $G_C$ are shown in Fig.~\ref{fig:GE}
and Fig.~\ref{fig:GC}, respectively. In both cases we plot the experimentally measured quantities $R_{EM}$ and 
$R_{SM}$ (\ref{GT}) that are related to the form factor ratios, normalized to the 
magnetic form factor $G_M$. Here, again, the asymptotic distribution amplitudes tend to give a better 
description.

In the future, one should try to constrain the parameters in the distribution amplitudes by making a combined 
fit to the LCSR for the electromagnetic and weak form factors of the nucleon, including $\Delta$-resonance 
production etc. In order to make this program fully quantitative one first has to
calculate radiative corrections to the LCSRs, similar as this has been done for the pion 
form factor \cite{Braun:1999uj} and B-meson decays \cite{Bdecays}.
The corresponding task goes beyond the scope of this paper.

%
%%%%%%%%%%%%%%%%%%     FIGURE 6          %%%%%%%%%%%%%%%%%%%%%%%%%%%%
%%%%%%%%%%%%%%%%%%%%%%%%%%%%%%%%%%%%%%%%%%%%%%%%%%%%%%%%%%%%%%%%%%%%%%
%

%
%%%%%%%%%%%%%%%%%%     FIGURE 7          %%%%%%%%%%%%%%%%%%%%%%%%%%%%
%%%%%%%%%%%%%%%%%%%%%%%%%%%%%%%%%%%%%%%%%%%%%%%%%%%%%%%%%%%%%%%%%%%%%%
%

\setcounter{equation}{0} \section{Summary and conclusions}

In this paper, we incorporated light-cone QCD sum rules approach to
calculate the purely nonperturbative soft contribution to the $\gamma^* p
\to \Delta^+$ transition. The soft  contribution corresponds to the
so-called Feynman mechanism of the large momentum transfer, 
which  is suppressed asymptotically by power(s) of $1/Q^2$ compared to the 
perturbative contribution of hard rescattering. We argued from the  very beginning that,
due to two hard gluon exchanges required in the pQCD contribution,
the latter is  numerically suppressed by a factor of 1/100
compared to the soft term.  Indeed, our results for the dominant magnetic
form factor $G_M (Q^2)$ are rather close to the experimental data in the
region above $Q^2 \sim 2$ GeV$^2$.  This confirms the general wisdom that 
the soft (end-point) contribution is dominant at the experimentally accessible 
momentum transfers. Moreover, the inspection shows that the soft contribution 
is dominated by quark configurations involving a ``minus'' light-cone projections 
of one of the quark field operators, which can be reinterpreted as importance 
of the orbital angular momentum (cf.~\cite{Ji:2002xn,Ji:2003yj}).

In the region of low $Q^2 < 2$~GeV$^2$ our results for the magnetic form factor
appear to be factor two below the data. One reason for that are most likely the 
so-called ``bilocal power corrections'' that correspond to
long-distance propagation in the $Q^2$ channel.     
Such terms were studied in case of the pion 
\cite{Nesterenko:1984tk} and nucleon \cite{Belyaev:1992xf}
form factors,
and they provided a sizable enhancement  of form factors at low momentum transfers 
as compared to the extrapolation of the large-$Q^2$ results.
Another possibility is that the interpolating current for the $\Delta$-particle
is not good enough and couples strongly to the excited states.   
This can be checked by applying the light-cone QCD sum rules 
to the correlation function $\langle \Delta | J \eta_N |0 \rangle$ of the 
electromagnetic current and a local current $\eta_N$ with nucleon quantum numbers,
while the $\Delta$ is left explicitly in the final state $\langle \Delta |$.
To pursue such a program, one needs a systematic analysis of the distribution amplitudes
of the $\Delta$ as the first step, the task  which is interesting in its own. 

Our calculations  are in agreement with the experimental observation 
that  the  ratios $E2/M1$ and $C2/M1$ are small. The charge form factor $G_E$ 
appears to be very sensitive to the shape of the nucleon distribution amplitudes
so that the experimental data can easily be fitted by assuming moderate corrections to the asymptotic 
form, of the same sign but much smaller as compared to the CZ model.
However, 
both the $G_{E}(Q^2)$ and $G_{C}(Q^2)$ form factors in our calculations
appear as  small differences between expressions dominated 
by  the much larger form factor $G_{M}(Q^2)$, a situation  similar to that
encountered in the approach \cite{Belyaev:1995ya} based on the 
analysis of the 3-point correlator. For this reason, at this stage we restrict ourselves
to a conservative statement that  $G_{E}(Q^2)$ and $G_{C}(Q^2)$ 
are small compared to $G_{M}(Q^2)$ without insisting on a specific curve
(or even sign) for them.

 The inclusion of QCD radiative corrections  to the sum rules presents another important issue, 
 and is needed to make the theoretical studies of the $\gamma^* p \to
\Delta^+$ transition fully quantitative.  Such corrections contain both
 purely soft contributions and also  hard gluon exchanges,
 with nontrivial interplay and connections between these two types of
 dynamics. For example, for the pion form factor it was found \cite{Braun:1999uj} that 
 there exists  a partial cancellation between soft contributions and hard contributions of higher 
 twist. A big advantage of the light-cone sum rule technique is that it is free from 
 double counting: soft and hard contributions can be separated rigorously. 
 In the present case, such a separation becomes necessary starting at two-loop 
 $O(\alpha_s^2)$ corrections to the sum rules.  

To summarize, we believe that the light-cone sum rule approach currently offers  
the best compromise between theoretical rigor and the applicability to present and planned 
experiments involving elastic and transition form factors for baryons. This approach 
is rigorous as far as the separation of hard and soft dynamics is concerned, and provides
one with a useful tool for the study of the transition region between hard perturbative and soft 
nonperturbative QCD dynamics. One goal of such studies is to determine nucleon 
distribution amplitudes from the data on form factors, similar as parton distributions 
are extracted from the measured deep inelastic structure functions. Our work presents
a step in this direction.

{\bf Acknowledgements}

We would like to thank Paul Stoler and Sabit Kamalov for providing us with detailed tables of the 
experimental data. The work of  A.R. was supported by the US 
 Department of Energy  contract
DE-AC05-84ER40150 under which the Southeastern
Universities Research Association (SURA)
operates the Thomas Jefferson Accelerator Facility, 
and by the Alexander von Humboldt Foundation.

\appendix  
\renewcommand{\theequation}{\Alph{section}.\arabic{equation}}  
%\setcounter{table}{0}  
%\renewcommand{\thetable}{\Alph{table}}  
   
%\section*{Appendicies}  

\setcounter{equation}{0} \section{Lorentz-invariant decomposition}  
\label{app:a}  
\setcounter{equation}{0}  
\setcounter{table}{0} 

The correlation function $T_{\mu\nu}(P,q)$ has to satisfy two conditions 
\beq{cc}
  \gamma^\mu T_{\mu\nu}(P,q) =0\,,\qquad q^\nu T_{\mu\nu}(P,q) =0\,,  
\eeq
that follow from the property of the current $\eta_\mu$ and electromagnetic current conservation, respectively. 
In addition, $T_{\mu\nu}(P,q)$ has to be proportional to the nucleon spinor. This suggests to define
\beq{lor}
   T_{\mu\nu}(P,q) = \sum_{k=1}^N T_k(P'^2,q^2) L^{(k)}_{\mu\nu}(P,q) \gamma_5 N(P) 
\eeq
 where $T_k(P'^2,q^2)$ are scalar functions in front of the Lorentz structures $L^{(k)}_{\mu\nu}(P',q)$ that 
satisfy the conditions in \Gl{cc} themselves, i.e. $\gamma^\mu L^{(k)}_{\mu\nu}(P',q)N(P)=0$ and 
 $q^\nu L^{(k)}_{\mu\nu}(P',q)N(P)=0$. 
The complete basis contains ten different Lorentz structures which can be chosen as
\bea{L}
L_{\mu\nu}^{(1)}&=&2(q_{\mu}\gamma_{\nu}q\s-g_{\mu\nu}q^{2})+\gamma_{\mu}(\gamma_{\nu}q^{2}-q_{\nu}q\s)\,,
\nonumber\\
L_{\mu\nu}^{(2)}&=&4(q_{\mu}P'_{\nu}-g_{\mu\nu}P'q)+\gamma_{\mu}(\gamma_{\nu}P'q-P'_{\nu}q\s)\,,
\nonumber\\ 
L_{\mu\nu}^{(3)}&=&P'_{\mu}(\gamma_{\nu}q\s-q_{\nu})-\frac14m_{P}(\gamma_{\mu}q\s\gamma_{\nu}
   -q_{\nu}\gamma_{\mu})+\frac14\gamma_{\mu}(\gamma_{\nu}q^{2}-q_{\nu}q\s)
+\frac12\gamma_{\mu}(\gamma_{\nu}P'q-P'_{\nu}q\s)\,,
\nonumber\\ 
L_{\mu\nu}^{(4)}&=&P'qP'_{\mu}\gamma_{\nu}q\s-P'_{\mu}P'_{\nu}q^{2}
 +m_{P}(P'qP'_{\mu}\gamma_{\nu}-P'_{\mu}P'_{\nu}q\s)+P'^{2}(q_{\mu}P'_{\nu}-g_{\mu\nu}P'q)\,,
\nonumber\\ 
L_{\mu\nu}^{(5)}&=& q_{\mu}q_{\nu}-g_{\mu\nu}q^{2}
+\frac14\gamma_{\mu}(\gamma_{\nu}q^{2}-q_{\nu}q\s) 
\eea
and
\bea{O}
L_{\mu\nu}^{(6)}&=&2(q_{\mu}\gamma_{\nu}-g_{\mu\nu}q\s)-(\gamma_{\mu}q\s\gamma_{\nu}-q_{\nu}\gamma_{\mu})\,,
\nonumber\\
L_{\mu\nu}^{(7)}&=&4(q_{\mu}P'_{\nu}q\s-g_{\mu\nu}P'qq\s)
      +2P'qq_{\nu}\gamma_{\mu}-P'q\gamma_{\mu}q\s\gamma_{\nu}-P'_{\nu}\gamma_{\mu}q^{2}\,,
\nonumber\\
L_{\mu\nu}^{(8)}&=&P'_{\mu}(\gamma_{\nu}q^{2}-q_{\nu}q\s)
   +P'_{\mu}m_{P}(\gamma_{\nu}q\s-q_{\nu})
   +\frac12m_{P}(\gamma_{\mu}\gamma_{\nu}P'q-\gamma_{\mu}P'_{\nu}q\s)
-\frac14P'^{2}(\gamma_{\mu}q\s\gamma_{\nu} -q_{\nu}\gamma_{\mu})
\nonumber\\&&
+\frac12(2P'q\gamma_{\mu}q_{\nu}
-\gamma_{\mu}P'qq\s\gamma_{\nu}-P'_{\nu}q^{2}\gamma_{\mu})\,,
\nonumber\\
L_{\mu\nu}^{(9)}&=&q_{\mu}q_{\nu}q\s-g_{\mu\nu}q^{2}q\s
   -\frac14q^{2}(\gamma_{\mu}q\s\gamma_{\nu}-\gamma_{\mu}q_{\nu})\,,
\\
L_{\mu\nu}^{(10)}&=&m_{P}P'_{\mu}(\gamma_{\nu}q\s-q_{\nu})
   +P'_{\mu}(\gamma_{\nu}q^{2}-q_{\nu}q\s)+2(P'qP'_{\mu}\gamma_{\nu}
-P'_{\mu}P'_{\nu}q\s)-\frac14 P'^{2}(\gamma_{\mu}q\s\gamma_{\nu}-q_{\nu}\gamma_{\mu})\,,
\nonumber 
\eea
where the two groups roughly correspond to contributions with even and odd number of gamma-matrices, 
respectively.

Taking the necessary light-cone projections we obtain for the amplitudes in (\ref{decomp}):
\bea{relate22}
 {\cal A} &=& Q^2\, T_4\,,
\nonumber\\
  {\cal B} &=& -m_P\, T_4 -2\, T_{10}\,,
\nonumber\\
  m_P {\cal C} &=& -Q^2[m_P\, T_4+2\,T_7- T_8+2\, T_{10}]\,,
\nonumber\\
   m_P {\cal D} &=&-2\,T_2+T_3+(2P'q-m_P^2)\,T_4+m_P\,T_8\,,
\nonumber\\
{\cal E} &=& m_P(P'q)\,T_4-Q^2\,T_8+(2P'q-Q^2)\,T_{10}\,,
\nonumber\\
{\cal F} &=& T_3+ (P'q)\,T_4+m_P\,T_8+m_P\,T_{10}\,,
\nonumber\\
{\cal G} &=& Q^2\,T_7+\frac12 Q^2\,T_8\,,
\nonumber\\
{\cal H} &=& -T_2-\frac12 T_3 -\frac12 m_P\,T_8\,,
\nonumber\\
{\cal I} &=&-Q^2[3\,T_2-\frac12 T_3-(2P'q-m_P^2)\,T_4-\frac12 m_P\,T_8]\,,
\nonumber\\
{\cal J} &=&- Q^2[ m_P\,T_4+3T_7-\frac12 T_8+2T_{10}]\,.
\eea 
Note that four invariant amplitudes $T_1$, $T_5$, $T_6$, $T_9$ do not contribute to the 
next-to-leading twist accuracy. This implies that the invariant functions in (\ref{decomp})
obey four relations which can be chosen as in Eqs.~(\ref{RScond}) and (\ref{LorSym}) in the text.  

\setcounter{equation}{0} \section{Nucleon Distribution Amplitudes}  
\label{app:b}  
\setcounter{equation}{0}  
\setcounter{table}{0}  

To make the paper self-contained, we collect here the necessary information on the nucleon
distribution amplitudes that enter the sum rules. Our presentation follows Ref.~\cite{Braun:2000kw}.

A standard tool to study hadron distribution amplitudes is  
to use constraints on operator mixing and equation of motion relations that originate  
from the conformal symmetry of the QCD Lagrangian \cite{Braun:2003rp}.
This approach  suggests an expansion in contributions of increasing conformal spin.
The leading-order contributions with the lowest conformal spin are 
usually referred to as asymptotic distribution amplitudes. To this accuracy there 
are only three nonperturbative parameters involved. One obtains~\cite{Braun:2000kw}
{asymptotic twist-3 distribution amplitudes:}
\bea{Asy-twist-3} 
V_1(x_i) = 120 \, x_1 x_2 x_3 f_N\, , \, \, \,
&& 
T_1(x_i) =  120 \, x_1 x_2 x_3 f_N \,,
\eea
{asymptotic twist-4 distribution amplitudes}
\bea{Asy-twist-4} 
V_2(x_i)  &=& 
2 \, x_1 x_2 \left[ 5 (1 + x_3) f_N + 6 \lambda_1  \right] \,,
%\hspace{0.5cm}
%A_2(x_i)  = 30 \, x_1 x_2 (x_1 - x_2) f_N \,,
\hspace*{1cm}
T_2(x_i) = 4 \,x_1 x_2 \lambda_2 \,,
\nn \\
V_3(x_i)  &=& 
x_3 \left[
5 \left(1 + 2 x_1 x_2 - 4 x_3 + 3 (x_1^2 + x_2^2 + x_3^2) \right) f_N
                   - 6 (1 - x_3) \lambda_1 \right]
\,, 
%\nn \\
%A_3(x_i) &=& - 2 (x_1-x_2) x_3 
%\left[5 (2 - 3 x_3) f_N - 3 \lambda_1
%\right] \, ,
\nn \\
T_3(x_i)  &=& x_3 \left[
5 \left(1 + 2 x_1 x_2 + 2 x_3 - 3 \left(x_1^2 + x_2^2 + x_3^2 \right) \right) f_N
                  + (1 - x_3) \lambda_2\right]
\, , \nn \\
T_7(x_i)  &=& x_3 \left[
5 \left(1 + 2 x_1 x_2 + 2 x_3 - 3 \left(x_1^2 + x_2^2 + x_3^2 \right) \right) f_N
                  - (1 - x_3) \lambda_2
\right]
\,,
\eea
{asymptotic twist-5 distribution amplitudes:}
\bea{Asy-twist-5} 
V_4(x_i) &=& 
- \frac{1}{3} \left[28 - 65 (x_1^2 + x_2^2)- 30 x_1 x_2 - 13 x_3 - 15 x_3^2 \right] f_N 
- \left[ 1 + (x_1- x_2)^2 - x_3^2 \right] \lambda_1
\,,
%\nn\\
%A_4(x_i) &=& \frac{1}{3} (x_1 -x_2)
%\left[ (37 - 80 x_3) f_N -6 \lambda_1 \right]\,,
\nn \\
T_4(x_i) &=& - \frac{1}{3} 
\left[28 - 65 (x_1^2 + x_2^2) + 30 x_1 x_2 - 43 x_3 + 15 x_3^2\right] f_N  
\nonumber\\&&{}
+\frac16 \left[1 + x_1^2 - 6 x_1 x_2 + x_2^2 + 2 x_3 - 3 x_3^2\right] \lambda_2\,,
\nn \\
T_8(x_i) &=& 
- \frac{1}{3} \left[28 - 65 (x_1^2 + x_2^2) + 30 x_1 x_2 - 43 x_3 + 15 x_3^2\right] f_N 
\nn \\
&&{}-\frac16 \left[1 + x_1^2 - 6 x_1 x_2 + x_2^2 + 2 x_3 - 3 x_3^2\right] \lambda_2 
\,,
\nn \\
V_5(x_i) &=& - \frac{2}{3} x_3 (28 - 65 x_3) f_N + 2 x_3 (1 + x_3) \lambda_1\,, 
\hspace{0.5cm}
%A_5(x_i) = 2 (x_1-x_2) x_3 \left(5 f_N - \lambda_1 \right)
T_5(x_i) \,  = \, \frac{2}{3} x_3 (1 + x_3) \lambda_2 \,,
\eea
{asymptotic twist-6 distribution amplitudes:}
\bea{Asy-twist-6} 
V_6(x_i) &=& \frac{1}{3} (5 + 3 x_3) f_N - \frac{2}{5} (1 - 3 x_3) \lambda_1 \,,
\hspace{0.5cm}
%A_6(x_i) = - \frac{1}{5} (x_1 - x_2) (5 f_N - 2 \lambda_1) \,,
%\nn \\
T_6(x_i) = \frac{1}{3} (8 - 6 x_3) f_N\,,
\eea
{and, finally, the $x^2$-corrections to the same accuracy:}
\bea{Asy-x2} 
V_1^{M(d)}(x_3) &=& \frac{x_3^2}{24} \left[
(1 - x_3) \left( (215 - 529 x_3 + 427 x_3^2 - 109 x_3^3) + 4 \ln[x_3] \right) f_N 
+ 16 (1 - x_3)^3 \lambda_1 
\right]
\, ,\nn \\
%V_1^{M(u)}(x) &=& \frac{x^2}{72} (1 - x)^3 \left[(565 - 417 x) f_N - 24 \lambda_1 \right]
%\, ,\nn \\
%A_1^{M(u)}(x) &=& \frac{x^2}{72} (1 - x)^3 \left[(43 + 105 x) f_N  - 24 \lambda_1 \right]
%\nn \\
T_1^{M(d)}(x_3) &=& \frac{x_3^2}{9} (1 - x_3)^3 \left[(76 - 39x_3) f_N - 6 \lambda_1 \right]\,,
%\nn \\
%T_1^{M(u)}(x) &=& \frac{x^2}{48}
%\left[((1 - x)(389 - 1051 x + 949 x^2 - 283 x^3) + 4 \ln[x]) f_N
%+
%16(1 - x)^3 \lambda_1
%\right]
%\nn \\
\eea 
where 
\bea{VTM}
{\cal V}_1^M(Pz) &:=& \int_0^1 dx_3 \; e^{-i x_3 (P\cdot z)}\, V_1^{M(d)}(x_3)\,,
\nonumber\\
{\cal T}_1^M(Pz) &:=& \int_0^1 dx_3 \; e^{-i x_3 (P\cdot z)}\, T_1^{M(d)}(x_3)\,,
\eea
cf. (\ref{zerl}), (\ref{fourier}).

The nonperturbative parameters $f_N$, $\lambda_1$ and $\lambda_2$ correspond to 
the vacuum-to-nucleon matrix elements of the three-quark local operators \cite{Braun:2000kw}.
The existing estimates come from QCD sum rules. At the scale $\mu = 1$ GeV one gets
\bea{numeric}
f_N  &=&    (5.0 \pm 0.3) \cdot 10^{-3} \mbox{GeV}^2 \; 
\cite{King:1986wi,Chernyak:1987nu}\,,
\nonumber\\
\lambda_1  &=&  - (2.7 \pm 0.5) \cdot 10^{-2} \mbox{GeV}^2 \; \cite{Braun:2000kw,Lenz05}\,,
\nonumber \\
\lambda_2  &=&    (5.4 \pm 1.0) \cdot 10^{-2} \mbox{GeV}^2 \; \cite{Braun:2000kw,Lenz05}\,.
\eea
Calculations in this work are done also using a more sophisticated model for the 
distribution amplitudes, taking into account corrections that correspond to the 
contributions of operators with the next-to-leading conformal spin \cite{Braun:2003rp}.
The corresponding expressions are given in \cite{Braun:2000kw} (see also \cite{Lenz05}). They include five more 
parameters (the first two leading twist, and the rest twist-4) which we choose as
\bea{corrections}
&& A_1^u  =    0.38 \pm 0.15  \; \cite{Chernyak:1984bm}\,,
\nn \\
&& V_1^d  =    0.23 \pm 0.03  \; \cite{Chernyak:1984bm}\,,
\nn\\
&& f_1^d  =    0.40 \pm 0.05  \; \cite{Braun:2000kw,Lenz05}\,,
\nn \\
&& f_2^d  =    0.22 \pm 0.05  \; \cite{Braun:2000kw,Lenz05}\,,
\nn\\
&& f_1^u  =    0.07 \pm 0.05  \; \cite{Braun:2000kw,Lenz05}\,.
\eea  
{}For a comparison, the asymptotic distribution amplitudes correspond in this
more general parametrization to $A_1^u =0$, $V_1^d  = 1/3$,   $f_1^d = 3/10$, 
$f_2^d = 4/15$ and  $f_1^u = 1/10$, see \cite{Braun:2000kw,Lenz05} for details.

Note that there is a mismatch between the twist classification 
of the distribution amplitudes that implies counting powers of the 
large momentum $p_+$, and the twist classification of the  
local operator matrix elements: E.g. the parameters of the 
twist-4 distribution amplitudes depend both on the 
the leading twist-3 matrix element $f_N$ and the 
twist-4 matrix elements $\lambda_1,\lambda_2$. Such 
contributions of lower-twist matrix elements to higher-twist 
distribution amplitudes are
well known and usually referred to as Wandzura-Wilczek contributions.
In particular the distribution amplitudes  of twist-5 and twist-6 are 
entirely of Wandzura-Wilczek type since there exist no geniune 
twist-5 and twist-6 operators to this order in the conformal expansion.

\setcounter{equation}{0} \section{QCD sum rule for $\Delta$-resonance}  
\label{app:c}  
\setcounter{equation}{0}  
\setcounter{table}{0}  

The coupling $\lambda_\Delta$ (\ref{lambdaD}) can be found from the Belyaev-Ioffe
sum rule \cite{Belyaev:1982sa} for the correlation function of the two $\eta_\mu$ currents.
The sum rule reads
\bea{BI-SR}
 \frac13 {\lambda_\Delta^2} \, e^{-m_\Delta^2/M^2} &=& 
 \frac15 M^6\left[1-e^{-s_0/M^2}\left(1+\frac{s_0}{M^2}+ \frac12 \frac{s_0^2}{M^4}
 \right)\right]
    -\frac{5b}{72} M^2 \Big(1- e^{-s_0/M^2}\Big) 
\nonumber\\&&{}
+ \frac{4a^2}{3} 
    -\frac{7m_0^2a^2}{9 m_B^2}, 
\eea
where $M^2$ is the Borel parameter, $s_0$ the continuum threshold and
\bea{condensates}
    a &=& -(2\pi)^2 \left\langle \bar q q \right\rangle = 0.55~\mbox{GeV}^3\,, 
\nonumber\\
    b &=& (2\pi)^2 \left\langle \frac{\alpha_s}{\pi} G^2 \right\rangle = 0.47~\mbox{GeV}^4\,,
\nonumber\\
   m^2_0 &=& 0.8~\mbox{GeV}^2\,.
\eea
Using an experimental value $m_\Delta=1232$~MeV we found that the sum rule (\ref{BI-SR}) 
is stable in a broad interval of Borel parameters $M^2 > 1.5$~GeV$^2$ 
for the choice of the continuum threshold 
$s_0 =3.0\pm 0.2$~GeV$^2$. The corresponding value of the coupling is 
equal to $\lambda_\Delta = 2.0\pm 0.2$~GeV$^3$.

 %%%%%%%%%%%%%%%%%%%%%%%%%%%%%%%%%%%%%%%%%%%%%%%%%%%%%%%%%%%%%%%%%%%%

\end{document}